\documentclass[letterpaper,journal,twocolumn]{IEEEtran}
\usepackage{tikz}
\usetikzlibrary{positioning,fit,calc}
\usetikzlibrary{decorations.pathreplacing,calligraphy}
\usepackage{geometry}
\usepackage{cite}
\usepackage{array}
\usepackage{bm}
\usepackage{bbm}
\usepackage{amssymb}
\usepackage{standalone}
\usepackage{setspace}
\usepackage{amsmath}
\usepackage{graphicx}
\usepackage{float}
\usepackage{booktabs}
\usepackage{multirow}
\usepackage{epsfig}
\usepackage[caption=false,font=footnotesize]{subfig}
\usepackage[nolist]{acronym}
\usepackage[implicit=false]{hyperref}
\hypersetup{hidelinks,
	colorlinks=true,
	allcolors=black,
	pdfstartview=Fit,
	breaklinks=true}

\definecolor{Red}{RGB}{246,113,72}
\definecolor{Grn}{RGB}{97,191,240}
\definecolor{Mred}{RGB}{217,83,25}
\definecolor{Mgrn}{RGB}{119,173,48}
\definecolor{Mblu}{RGB}{0,114,190}

\newtheorem{proposition}{Proposition}

\newtheorem{theorem}{Theorem}
\newtheorem{lemma}{Lemma}
\newtheorem{remark}{Remark}

\newcommand{\UpRoman}[1]{\MakeUppercase{\romannumeral #1}}
\newcommand{\SNR}{{\rm SNR}}

\begin{document}

\title{Probabilistic Shaped Multilevel Polar Coding for Wiretap Channel}
\author{Li Shen, 
        Yongpeng Wu, \IEEEmembership{Senior Member, IEEE},
        Peihong Yuan, \IEEEmembership{Member, IEEE},
        Chengshan Xiao, \IEEEmembership{Fellow, IEEE},
        Xiang-Gen Xia, \IEEEmembership{Fellow, IEEE},
        and Wenjun Zhang, \IEEEmembership{Fellow, IEEE}


\thanks{Li Shen, Yongpeng Wu, and Wenjun Zhang are with the Department of Electronic Engineering, Shanghai Jiao Tong University, Shanghai 200240, China (e-mail: \href{mailto:shen-l@sjtu.edu.cn}{shen-l@sjtu.edu.cn}; \href{mailto:yongpeng.wu@sjtu.edu.cn}{yongpeng.wu@sjtu.edu.cn}; \href{mailto:zhangwenjun@sjtu.edu.cn}{\mbox{zhangwenjun@sjtu.edu.cn}}).}

\thanks{Peihong Yuan is with the Research Laboratory of Electronics (RLE), Massachusetts Institute of Technology (MIT), Cambridge, MA, USA (email: \href{mailto:phyuan@mit.edu}{phyuan@mit.edu}).}

\thanks{Chengshan Xiao is with the Department of Electrical and Computer Engineering, Lehigh University, Bethlehem, PA 18015, USA (e-mail: \href{mailto:xiaoc@lehigh.edu}{\mbox{xiaoc@lehigh.edu}}).}

\thanks{Xiang-Gen Xia is with the Department of Electrical and Computer Engineering, University of Delaware, Newark, DE 19716, USA (e-mail: \href{mailto:xianggen@udel.edu}{\mbox{xianggen@udel.edu}}).}
}

\begin{acronym}
\acro{ASK}{amplitude shift keying}
\acro{AWGN}{additive white Gaussian noise}
\acro{B-DMC}{binary-input \ac{DMC}}
\acro{BER}{bit error rate}
\acro{BICM}{bit-interleaved coded modulation}
\acro{CRC}{cyclic redundancy check}
\acro{DM}{distribution matcher}
\acro{DMC}{discrete memoryless channel}
\acro{DMWC}{discrete memoryless wiretap channel}
\acro{FEC}{forward error correction}
\acro{GS}{geometric shaping}
\acro{HY}{Honda-Yamamoto}
\acro{i.i.d.}{independent and identically distributed}
\acro{LDPC}{low-density parity-check}
\acro{LLR}{log-likelihood ratio}
\acro{MAP}{maximum a posteriori}
\acro{MB}{Maxwell-Boltzmann}
\acro{MI}{mutual information}
\acro{MLC}{multilevel coding}
\acro{MSD}{multistage decoding}
\acro{PAS}{probabilistic amplitude shaping}
\acro{PLS}{physical layer security}
\acro{PS}{probabilistic shaping}
\acro{PSK}{phase shift keying}
\acro{QAM}{quadrature amplitude modulation}
\acro{RF}{Rayleigh fading}
\acro{SC}{successive cancellation}
\acro{SCL}{successive cancellation list}
\acro{SNR}{signal-to-noise ratio}
\end{acronym}

\renewcommand{\IEEEQED}{\IEEEQEDopen}

\maketitle
\pagestyle{empty}
\thispagestyle{empty}
\begin{abstract}
    A wiretap channel is served as the fundamental model of physical layer security techniques, where the secrecy capacity of the Gaussian wiretap channel is proven to be achieved by Gaussian input. However, there remains a gap between the Gaussian secrecy capacity and the secrecy rate with conventional uniformly distributed discrete constellation input, e.g. amplitude shift keying (ASK) and quadrature amplitude modulation (QAM). In this paper, we propose a probabilistic shaped multilevel polar coding scheme to bridge the gap. Specifically, the input distribution optimization problem for maximizing the secrecy rate with ASK/QAM input is solved. Numerical results show that the resulting sub-optimal solution can still approach the Gaussian secrecy capacity. Then, we investigate the polarization of multilevel polar codes for the asymmetric discrete memoryless wiretap channel, and thus propose a multilevel polar coding scheme integration with probabilistic shaping. It is proved that the scheme can achieve the secrecy capacity of the Gaussian wiretap channel with discrete constellation input, and satisfies the reliability condition and weak security condition. A security-oriented polar code construction method to natively satisfies the leakage-based security condition is also investigated. Simulation results show that the proposed scheme achieves more efficient and secure transmission than the uniform constellation input case over both the Gaussian wiretap channel and the Rayleigh fading wiretap channel.
\end{abstract}

\begin{IEEEkeywords}
Physical layer security, asymmetric wiretap channels, multilevel coding, polar codes, probabilistic shaping.
\end{IEEEkeywords}

\section{Introduction}
\IEEEPARstart{A}{s} a complement or an alternative to classical cryptographic techniques, the \ac{PLS} techniques provide an additional layer of security for future wireless communication networks by exploiting the inherent randomness of wireless channels \cite{Irram2022Physical}.
The wiretap channel model was first proposed by Wyner \cite{wyner1975wire} in 1975 and has become the fundamental scenario in \ac{PLS} researches. Wyner 
showed that a secure transmission with positive rate is possible by applying appropriate coding and signal processing scheme, and derived the secrecy capacity of the \ac{DMWC}. In \cite{leung1978gaussian}, the authors extended the concept of secrecy capacity to the Gaussian wiretap channel, and proved that the secrecy capacity is achieved by the Gaussian codebook. Under finite complex constellation input constraints, the secrecy capacity of the Gaussian wiretap channel was also studied in \cite{raghava2010secrecy}. Moreover, the secrecy capacity of the fading wiretap channel has been considered \cite{Rezki2014secrecy,Lin2016fast}.

In addition to the information theoretic research on secrecy capacity introduced above, some practical wiretap codes are required to guarantee secure communication. The work in \cite{Thangaraj2007appli} presented a wiretap coding scheme based on \ac{LDPC} codes and shown that it achieves secrecy capacity in some cases. In \cite{mahdavifar2011achieving}, the authors proposed a secrecy capacity achieving polar coding scheme for the binary-input symmetric \ac{DMWC} and proved that the scheme satisfies the reliability condition and the weak security condition. To avoid the measure of information theoretic secrecy metric like the security condition, a \ac{BER} based secrecy metric named security gap was suggested in \cite{klinc2011ldpc} and has become one of the criteria for wiretap code design \cite{klinc2011ldpc,Baldi2012coding,Zhang2014polar,Nooraiepour2017Rand,Pfeiffer2022design}. The secrecy metrics for fading wiretap channel were investigated in \cite{Baldi2019physical}. Moreover, with the popularity of artificial intelligence technologies in recent years, the wiretap code design via autoencoder has attracted the attention of researchers \cite{Besser2020wiretap,Rana2023short}. Furthermore, to achieve a higher transmission rate, it is necessary to use a high-order discrete modulation. For instance, \cite{Matsumine2022Security} optimized bit-labeling of \ac{BICM} systems for the Gaussian wiretap channel, and \cite{Pfeiffer2022multi} proposed a \ac{MLC} scheme based on the punctured LDPC codes for physical layer security.

Nevertheless, for the Gaussian wiretap channel with standard uniformly distributed constellation input, such as conventional \ac{ASK} constellations and \ac{QAM} constellations, there remains a gap between the achievable secrecy rate and the Gaussian secrecy capacity \cite{raghava2010secrecy}, and none of the existing works have considered bridging the gap.

Inspired by constellation shaping techniques, such as \ac{GS} and \ac{PS}, which are commonly performed in point-to-point communication systems to achieve a Gaussian-like input distribution, we may leverage constellation shaping to narrow the gap. The \ac{GS} designs non-equidistant constellations with uniformly distributed constellation points, e.g., constellations for ATSC 3.0 \cite{Loghin2016non}, but their performance is worse than \ac{PS} under the same modulation order \cite{Steiner2020Coding}. The \ac{PS}, on the other hand, generates constellation points from a standard constellation with non-uniform probability, e.g., trellis shaping \cite{Forney1992Trellis}, shell mapping \cite{Khandani1993Shaping}, \ac{PAS} \cite{bocherer2015bandwidth} and \ac{HY} scheme \cite{honda2013polar}. Particularly, \ac{PAS} and the \ac{HY} scheme can implement \ac{PS} with lower complexity and more flexible transmission rate. The \ac{PAS} scheme concatenates a \ac{DM} with a systematic \ac{FEC} code to perform \ac{PS} and channel coding \cite{bocherer2015bandwidth}, and is proved to achieve the capacity of \ac{AWGN} channel \cite{bocherer2023proba}. Alternatively, \cite{honda2013polar} revealed the polarization of polar codes for the binary-input asymmetric \ac{DMC}\footnote{For asymmetric settings, the capacity achieving input distribution of the codewords is not always uniform.} and thus proposed a polar coding scheme to achieve the asymmetric capacity. With \ac{HY} scheme, the \ac{PS} and \ac{FEC} can be jointly implemented by just performing a polar decoder. Furthermore, the \ac{HY} scheme has been extended to higher-order modulations by using the \ac{MLC} structure \cite{icscan2020sign, runge2022multilevel}.

However, it does not mean that constellation shaping can be directly applied for transmission over the Gaussian wiretap channel, since achieving the secrecy capacity not only requires an optimal input distribution, but also implies that the amount of information leaked to the eavesdropper should tend to zero. More specifically, although a desired distribution can be achieved through constellation shaping, many wiretap codes, like \cite{klinc2011ldpc,Baldi2012coding,Zhang2014polar,Nooraiepour2017Rand,Pfeiffer2022design,Matsumine2022Security,Pfeiffer2022multi}, are designed merely to impose high \ac{BER} at Eve, without providing any promise of information theoretic security. Moreover, some shaping schemes, e.g. \ac{GS} and \ac{PAS}, need to work with a \ac{BICM} structure and/or a \ac{DM}, which makes the analysis of information leakage challenging. Due to the above considerations, we turn to multilevel polar coding since polar codes have shown the ability to achieve the secrecy capacity of binary-input symmetric \ac{DMWC} \cite{mahdavifar2011achieving} and the capacity of constellation-input asymmetric \ac{DMC} \cite{runge2022multilevel}.

In this paper, we propose a probabilistic shaped multilevel polar coding scheme achieving the secrecy capacity of the Gaussian wiretap channel with discrete constellation input. We investigate the polarization for the asymmetric \ac{DMWC} with constellation input, leading to a random bit set and a shaping bit set in addition to the information bit set and frozen bit set in conventional polar codes. These two additional sets are considered separately in \cite{mahdavifar2011achieving} and \cite{runge2022multilevel}, respectively. Hence, we need to take into account the simultaneous introduction of the two sets and re-establish the conclusions on capacity achieving, reliability, and security, following the basic ideas in \cite{mahdavifar2011achieving} and \cite{runge2022multilevel}. Furthermore, since the secrecy capacity achieving input distribution is the one that maximizes the secrecy rate rather than the mutual information as in conventional shaping schemes, an input distribution optimization problem for the Gaussian wiretap channel should be solved. To enable secure transmission, a practical code construction method under security condition constraint should also be investigated. The main contributions of this paper are detailed as follows.

\begin{enumerate}
    \item We formulate the problem of optimizing the input distribution to maximize the secrecy rate of the Gaussian wiretap channel under discrete constellation input constraints, and simplify the problem by adopting the \ac{MB} distribution. Specifically, given an \ac{ASK}/\ac{QAM} constellation input, a sub-optimal solution can be found by traversing the feasible interval. Numerical results show that the sub-optimal solution achieves a secrecy rate close to the Gaussian secrecy capacity with negligible gap.
    \item We propose a probabilistic shaped multilevel polar coding scheme. For the asymmetric \ac{DMWC} with discrete constellation input, we show that the bit positions of a multilevel polar code will asymptotically polarize into four sets, which are placed with frozen bits, message bits, random bits and shaping bits, respectively. Such a polar code is proven to be secrecy capacity achieving and to satisfy the reliability condition and the weak security condition as code length goes to infinity. We also remark that the above conclusions apply to the degraded Gaussian wiretap channel.
    \item We propose a security-oriented polar code construction method for the proposed coding scheme at finite block length regimes, which natively satisfies the leakage-based security condition given a desired security gap. For completeness, the encoding and decoding procedures and complexity analysis are also provided.
    \item Simulation results show that the proposed scheme can achieve more efficient and secure transmission than the uniform input case over both the Gaussian wiretap channel and the Rayleigh fading (RF) wiretap channel, and the results corroborate the theoretical conclusions.
\end{enumerate}

The rest of the paper is organized as follows. Section \UpRoman{2} introduces some essential preliminaries. In Section \UpRoman{3}, we solve the input distribution optimization problem for maximizing the secrecy rate. The probabilistic shaped multilevel polar coding scheme is proposed in Section \UpRoman{4}. Section \UpRoman{5} introduces the coding procedure and code construction method. Section \UpRoman{6} presents our simulation results. Finally, Section \UpRoman{7} concludes this paper.

\textit{Notations:} Random variables are denoted by uppercase letters, e.g., $X$, while their realizations are denoted by lowercase letters, e.g., $x$. We denote $P_X$ and $E[X]$ by the probability distribution and expectation of $X$, respectively. Vectors are denoted by bold symbols, e.g., $\bm{X}$. The notations $H(X)$, $H(X|Y)$ and $I(X;Y)$ indicate the entropy of $X$, the entropy of $X$ conditioned on $Y$ and the \ac{MI} of $X$ and $Y$, respectively. The probability of an event is written as $\mathrm{Pr}\{\cdot\}$. Sets like alphabet are denoted by calligraphic letters, e.g., $\mathcal{X}$. $\mathcal{X}^C$ and $|\mathcal{X}|$ are the complement and cardinality of $\mathcal{X}$, respectively. The set difference is defined as $\mathcal{X} \setminus \mathcal{Y} = \mathcal{X} \cap \mathcal{Y}^C$. We denote the real field, positive integer set and binary field by $\mathbb{R}$, $\mathbb{N}$ and $\mathbb{F}_2$, respectively. The empty set is denoted by $\emptyset$. An index set $\{1, 2, \cdots, N\}$ is abbreviated as $[\![N]\!]$, and particularly $[\![0]\!]$ is equal to $\emptyset$. Given a vector $\bm{x}$ of length $N$ and $\mathcal{A} \subseteq [\![N]\!]$, we write $\bm{x}_\mathcal{A}$ to denote the sub-vector $[x_i]_{i\in \mathcal{A}}$. 


\section{Preliminaries}
\subsection{Wiretap Channel Model}
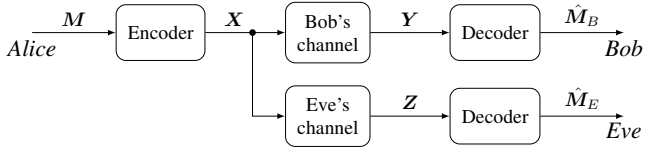
\begin{figure}[!tp]
    \setstretch{1}
    \centering
    \tikzset{
    box/.style ={
    rectangle, 
    rounded corners = 5pt, 
    minimum width = 1.5cm, 
    minimum height = 1cm, 
    inner sep = 6pt, 
    text centered,
    draw = black, 
    align = center,
    },
    arrow/.style ={
    -latex,
    },
}

\begin{tikzpicture}
\node[box] (Enc) at(0,0) {Encoder};
\node[box] (ChB) at([xshift=3cm]Enc) {Bob's \\ channel};
\node[box] (ChE) at([yshift=-1.5cm]ChB) {Eve's \\ channel};
\node[box] (DecB) at([xshift=3cm]ChB) {Decoder};
\node[box] (DecE) at([xshift=3cm]ChE) {Decoder};

\coordinate[label=below:{\large\textit{Alice}}] (Alice) at ([xshift=-1.5cm]Enc.west);
\coordinate[label=below:{\large\textit{Bob}}] (Bob) at ([xshift=1.5cm]DecB.east);
\coordinate[label=below:{\large\textit{Eve}}] (Eve) at ([xshift=1.5cm]DecE.east);

\draw[arrow] (Alice) --node[above]{$\bm{M}$} (Enc);
\draw[arrow] (Enc) --node[xshift=-0.2cm,above]{$\bm{X}$} (ChB);
\draw[arrow] ($(Enc)!0.55!(ChB)$) |- (ChE.west);
\draw[fill] ($(Enc)!0.55!(ChB)$) circle [radius=0.05];
\draw[arrow] (ChB) --node[above]{$\bm{Y}$} (DecB);
\draw[arrow] (ChE) --node[above]{$\bm{Z}$} (DecE);
\draw[arrow] (DecB) --node[above]{$\hat{\bm{M}}_B$} (Bob);
\draw[arrow] (DecE) --node[above]{$\hat{\bm{M}}_E$} (Eve);

\end{tikzpicture}
    \caption{The wiretap channel model.}
    \label{fig:wiretap}
\end{figure}

The well-known wiretap channel model \cite{wyner1975wire} is shown in Fig.~\ref{fig:wiretap}. A transmitter Alice wants to send a $K$-bit uniformly distributed message $\bm{M}$ to the legitimate receiver Bob, while the transmission is wiretapped by an eavesdropper Eve. The message $\bm{M}$ is encoded and mapped to a symbol sequence $\bm{X}$ of length $N$. Each element of $\bm{X}$ is taken from the constellation $\mathcal{X}$ of cardinality $Q=|\mathcal{X}|$. Bob and Eve receive $\bm{X}$ through their respective channels. The channel observations of Bob and Eve are denoted by $\bm{Y}$ and $\bm{Z}$, respectively. We assume that the coding scheme is perfectly known by both Bob and Eve. Thus, the observations $\bm{Y}$ and $\bm{Z}$ can be demapped and decoded to estimate $\hat{\bm{M}}_B$ and $\hat{\bm{M}}_E$, respectively.

\subsection{Secrecy Metrics} \label{Sec:PreSm}
\subsubsection{Reliability and Security}
In the wiretap channel model, Alice's transmission should achieve two objectives---reliability and security. Reliability means that Bob can reconstruct $\bm{M}$ with negligibly probability of error. Specifically, the reliability condition is given by
\begin{equation}
\label{eq:r_condt}
    \lim_{K\to\infty} \mathrm{Pr}\{\hat{\bm{M}}_B \neq \bm{M}\} = 0.
\end{equation}
Security refers to preventing Eve from extracting any information about $\bm{M}$. Usually, security is measured in terms of the number of bits leaked to Eve, also denoted as the \textit{information leakage} $I(\bm{M};\bm{Z})$. Specifically, the \textit{weak security} condition is defined as the leakage per coded symbol vanishes for infinite symbol length, i.e.,
\begin{equation}
\label{eq:ws_condt}
    \lim_{N\to\infty} \frac{1}{N} I(\bm{M};\bm{Z}) = 0
\end{equation}
while the \textit{strong security} condition \cite{Maurer1994} is defined as the leakage itself vanishes for infinite symbol length, i.e.,
\begin{equation}
\label{eq:ss_condt}
    \lim_{N\to\infty} I(\bm{M};\bm{Z}) = 0.
\end{equation}

\subsubsection{Secrecy Capacity}
The secrecy capacity defines the maximum theoretical transmission rate for reliable communication between Alice and Bob, while keeping Eve ignorant of the message. Specifically, for a degraded \ac{DMWC}, i.e., Eve's channel is a degraded version of Bob's channel, the achievable secrecy rate $R_s$ is calculated by 
\begin{equation}
\label{eq:Rs}
    R_s = I(X;Y)-I(X;Z)
\end{equation}
which is an upper bound on the transmission rate achieving both reliable and secure communication with a given input distribution $P_X$ of $X$. Then, the secrecy capacity is the maximum $R_s$ taken over all possible input distributions $P_X$ \cite{wyner1975wire,csiszar1978broadcast}, i.e.,
\begin{equation}
\label{eq:Cs}
    C_s = \max_{P_X} R_s = \max_{P_X} \left[ I(X;Y)-I(X;Z) \right].
\end{equation}

\subsubsection{Equivocation Rate}
As a measure of Eve's uncertainty about the message $\bm{M}$ from his observation $\bm{Z}$, the equivocation rate is calculated by
\begin{equation}
    R_e = \frac{1}{N} H(\bm{M}|\bm{Z}).
    \label{eq:Re}
\end{equation}
Let the transmission rate be $R_t = K/N$. A rate-equivocation pair $(\tilde{R}_t, \tilde{R}_e)$ is said to be achievable if, for all $\epsilon > 0$, there exists a code sequence of rate $R_t$ such that
\vspace{0 cm}
\begin{equation}
\begin{cases}
    \displaystyle \lim_{N \to \infty} R_t \geq \tilde{R}_t - \epsilon, \\
    \displaystyle \lim_{N \to \infty} R_e \geq \tilde{R}_e - \epsilon, \\
    \displaystyle \lim_{N \to \infty} \mathrm{Pr}\{\hat{\bm{M}}_B \neq \bm{M}\} \leq \epsilon.
\end{cases}
\end{equation}
The \textit{rate-equivocation region} is defined as the closure of all achievable rate-equivocation pairs $(\tilde{R}_t, \tilde{R}_e)$. Consider a degraded wiretap channel with a fixed input distribution $P_X(x)$, the achievable rate-equivocation region is given by \cite{wyner1975wire}
\begin{equation}
\begin{cases}
    \displaystyle \tilde{R}_e \leq \tilde{R}_t \leq I(X;Y), \\
    \displaystyle 0 \leq \tilde{R}_e \leq R_s.
\end{cases}
\label{eq:Re_rg}
\end{equation}

\subsubsection{Security Gap}
The secrecy metrics introduced above are information theoretic, which are difficult to compute in many practical coding scenarios and finite block length regimes. To this point, Klinc et al. proposed a practical secrecy metric named security gap that is measured by \ac{BER} \cite{klinc2011ldpc}. Specifically, let $P_e^B$ and $P_e^E$ denote the average \ac{BER} of Bob and Eve, respectively. Given a reliability threshold $P_{e,\max}^B$ and a security threshold $P_{e,\min}^E$, the reliability condition is said to be satisfied if $P_e^B \leq P_{e,\max}^B$, while the security condition is said to be achieved if $P_e^E \geq P_{e,\min}^E$. Assuming that Bob's and Eve's channels are corrupted by \ac{AWGN}, the security gap is given by
\begin{equation}
    \label{eq:Sg}
    S_g = \frac{\SNR_{B,\min}}{\SNR_{E,\max}}
\end{equation}
where $\SNR_{B,\min}$ denotes the minimum required \ac{SNR} of Bob's channel to achieve the reliability condition and $\SNR_{E,\max}$ denotes the maximum \ac{SNR} of Eve's channel to ensure the security condition. This metric defines how much the degradation of Eve's channel with respect to Bob's channel needs to be to achieve a secure transmission. Obviously, the security gap is expected to be as small as possible.

Moreover, when applying polar codes, the security can also be measured by the estimated information leakage. Benefiting from the special structure of polar codes, the amount of information leaked to Eve can be estimated by a modified \ac{SC} decoder \cite{taleb2021infor} or some polar construction methods \cite{bocherer2017efficient, tal2013construct}. Similar to the security condition in the form of \ac{BER}, the leakage-based security condition is defined as $L_k^E \leq L_{k,\max}^E$, where $L_k^E$ denotes the estimation of the normalized information leakage $I(\bm{M};\bm{Z})/K$ and $L_{k,\max}^E$ denotes the security threshold. In this case, the definition of the security gap still holds.

\subsection{Polar Codes}
Polar codes, proposed by Arıkan, are known as the first channel coding scheme proven to achieve the symmetric capacity of arbitrary \ac{B-DMC} \cite{arikan2009channel}. Let $W: X \to Y$ be a \ac{B-DMC} with input alphabet $\mathcal{X}=\{0,1\}$ and output alphabet $\mathcal{Y}$. Considering polar codes with block length $N = 2^n$ for $n \in \mathbb{N}$, the polar transform maps an input vector $\bm{u} \in \mathbb{F}_2^N$ to a codeword $\bm{x} \in \mathbb{F}_2^N$ by using
\begin{equation}
    \bm{x} = \bm{u}\bm{G}_N \quad\text{with}\quad \bm{G}_N=\bm{B}_N\bm{G}^{\otimes n}
\end{equation}
where $\bm{B}_N$ is the bit-reversal matrix as in \cite{arikan2009channel} and $\bm{G}^{\otimes n}$ denotes the $n$-th Kronecker power of $\bm{G} = \begin{bmatrix} 1 & 0 \\ 1 & 1 \end{bmatrix}$. The generator matrix $\bm{G}_N$ satisfies $\bm{G}_N^{-1} = \bm{G}_N$. Then, the codeword $\bm{x}$ is transmitted over $N$ independent uses of the \ac{B-DMC} $W$ and the received signal vector is $\bm{y} \in \mathcal{Y}^N$. With \ac{SC} decoding, the transition probability of the $i$-th sub-channel is given by
\begin{equation}
\label{eq:subch}
    W_N^{(i)}\left(\bm{u}_{[\![i-1]\!]}, \bm{y} | u_i \right) = P_{\bm{U}_{[\![i-1]\!]}, \bm{Y} | U_i} \left(\bm{u}_{[\![i-1]\!]}, \bm{y} | u_i \right).
\end{equation}

Arıkan showed that the bits in $\bm{u}$ asymptotically polarize into two sets \cite{arikan2009channel}:
\begin{align}
    \mathcal{G} &= \left\{ i \in [\![N]\!] : Z_B\left(U_i | \bm{U}_{[\![i-1]\!]}, \bm{Y} \right) < \delta_N \right\}, \\
    \mathcal{B} &= \left\{ i \in [\![N]\!] : Z_B\left(U_i | \bm{U}_{[\![i-1]\!]}, \bm{Y} \right) > 1-\delta_N \right\}
\end{align}
where $\delta_N \triangleq 2^{-N^\beta}$ for any $\beta \in \left(0,\frac{1}{2}\right)$ and $Z_B(\cdot|\cdot)$ denotes the Bhattacharyya parameter. More specifically, given that $(T,V) \sim P_{T,V}$ is an arbitrary pair of random variables, where $T \in \mathbb{F}_2$ and $V$ takes values in an arbitrary alphabet $\mathcal{V}$, the Bhattacharyya parameter \cite{arikan2010source} is defined as
\begin{equation}
\begin{aligned}
\label{eq:ZB}
    Z_B(T|V) &= 2\sum_{v\in\mathcal{V}} P_V(v) \sqrt{P_{T|V}(0|v) P_{T|V}(1|v)} \\
    &= 2\sum_{v\in\mathcal{V}} \sqrt{P_{T,V}(0,v) P_{T,V}(1,v)}
\end{aligned}
\end{equation}
and satisfies that $Z_B(T|V)$ is close to 0 or 1 if and only if $H(T|V)$ is close to 0 or 1 \cite[Proposition 2]{arikan2010source}, respectively. Therefore, the general idea of polar codes is to use the positions indexed by $\mathcal{G}$ for data transmission, while the remaining positions in $\bm{u}$ are filled with \textit{frozen} bits, which are set to fixed values known to both transmitter and receiver. As such, the data bit $U_i$, $i \in \mathcal{G}$, can be successively determined by $\bm{U}_{[\![i-1]\!]}$ and $\bm{Y}$ since $H\left(U_i | \bm{U}_{[\![i-1]\!]}, \bm{Y} \right) \approx 0$.


\section{Optimization of the Input Distribution for the Gaussian Wiretap Channel} \label{Sec:PxOpt}

We consider the Gaussian wiretap channel with \ac{ASK} input, where the channel input $X$ is taken from a $Q$-ary \ac{ASK} constellation $\mathcal{X}=\left\{\pm1, \pm3, \cdots, \pm(Q-1) \right\}$ with distribution $P_X$. Assuming that the transmit power is $P$, we scale $X$ by the constellation scaling factor $\Delta > 0$ to satisfy the power constraint. Thereby, the received signals at the legitimate receiver and the eavesdropper are given by 
\begin{align}
    Y&=\Delta X+N_B, \\
    Z&=\Delta X+N_E
\end{align}
respectively, where $N_B$ and $N_E$ follow the \ac{i.i.d.} complex Gaussian distribution with zero-mean and variances $\sigma_B^2$ and $\sigma_E^2$, i.e., $N_B \sim \mathcal{CN}(0,\sigma_B^2)$ and  $N_E \sim \mathcal{CN}(0,\sigma_E^2)$. We denote the \acp{SNR} of Bob's channel and Eve's channel by $\SNR_B=P/\sigma_B^2$ and $\SNR_E=P/\sigma_E^2$, respectively. If Eve’s channel is degraded with respect to Bob’s channel, we have $\SNR_B > \SNR_E$. 

To narrow the gap between the achievable secrecy rate with uniformly distributed \ac{ASK} input and the Gaussian secrecy capacity, we formulate the following input distribution optimization problem
\begin{equation}
\label{eq:Px_opt}
    \max_{P_X} ~R_s \qquad \text{s.t.} ~E\left[|\Delta X|^2\right] = P
\end{equation}
where $R_s$ is specifically calculated by (\ref{eq:Rs_calc}) at the top of next page.
\begin{figure*}[t]
    \begin{equation}
    \label{eq:Rs_calc}
    \begin{aligned}
        R_s = \sum_{x_i\in\mathcal{X}} P_X(x_i) & \left( E_{N_E} \left[ \log_2 \sum_{x_j\in\mathcal{X}} \frac{P_X(x_j)}{P_X(x_i)} \exp\left( -\frac{|n_E+\Delta(x_i-x_j)|^2 - |n_E|^2}{\sigma_E^2} \right) \right]\right. \\ 
        &\qquad\qquad\qquad \left. - E_{N_B} \left[ \log_2 \sum_{x_j\in\mathcal{X}} \frac{P_X(x_j)}{P_X(x_i)} \exp\left( -\frac{|n_B+\Delta(x_i-x_j)|^2 - |n_B|^2}{\sigma_B^2} \right) \right] \right).
    \end{aligned}
    \end{equation}
    \hrulefill
    \vspace*{4pt}
\end{figure*}
Since the secrecy capacity of the Gaussian wiretap channel is achieved by Gaussian input \cite{leung1978gaussian}, we use the \ac{MB} distribution family
\begin{equation}
    P_{X_\nu}(x) = \frac{e^{-\nu x^2}}{\sum_{x^\prime \in \mathcal{X}}e^{-\nu x^{\prime^2}}}, ~x\in \mathcal{X}
\end{equation}
with the parameter $\nu \geq 0$, which is a discrete Gaussian-like distribution that maximizes the entropy of the channel inputs subject to the power constraint \cite{kschischang1993optimal}. Thus, for a given scaling factor $\Delta$, we choose the input distribution
\begin{equation}
    P_{X_\Delta}(x)=P_{X_\nu}(x) \quad\text{with}\quad\nu: E\left[|\Delta X_\nu|^2\right] = P.
\end{equation}
As $E\left[|X_\nu|^2\right]$ is strictly monotonically decreasing in $\nu$ \cite{bocherer2015bandwidth}, the value of $\nu$ can be found efficiently by the \textit{bisection method}. Moreover, for $\nu = 0$, the value of $E\left[|X_\nu|^2\right]$ equals to $(Q^3-Q)/6$ and converges to $1$ for $\nu \to \infty$. As such, (\ref{eq:Px_opt}) simplifies to an optimization problem over $\Delta$ as follows
\begin{subequations}
\label{eq:Px_opt_s}
\begin{align}
    \max_{\Delta}\quad & R_s \\
    \text{s.t.}\quad & \sqrt{\frac{6P}{Q^3-Q}} \leq \Delta < \sqrt{P}, \label{eq:interval} \\
    & P_{X}(x) = P_{X_\Delta}(x).
\end{align}
\end{subequations}
We note that the optimization problem (\ref{eq:Px_opt_s}) differs from the optimization problem that maximizes mutual information $I(X;Y)$ in \cite{bocherer2015bandwidth} because $I(X;Y)$ is an unimodal function of $\Delta$, while $R_s$ is not (in fact, we find that $R_s$ is a bimodal function of $\Delta$ by numerical results). Hence, we propose to solve the optimization problem by traversing the feasible interval (\ref{eq:interval}) with a small step size and thus shape the input distribution to an \ac{MB} distribution that maximizes the achievable secrecy rate $R_s$. The resulting scaling factor and the corresponding input distribution are denoted by $\Delta^*$ and $P_{X^*}$, respectively. Note that the distribution $P_{X^*}$ is a sub-optimal solution to (\ref{eq:Px_opt}) in general, due to the choice of \ac{MB} distribution family and the way we traverse the interval (\ref{eq:interval}). However, the following simulations will show that the gain from further optimizing is negligible.

\begin{figure}[!t]
    \centering
    \includegraphics[width=0.48\textwidth]{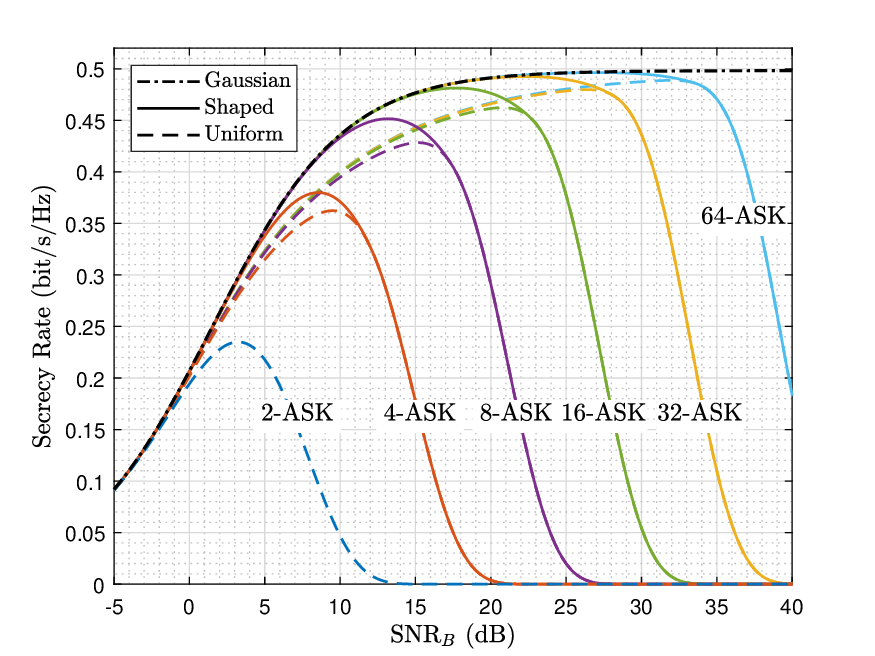}
    \caption{The achievable secrecy rate for the Gaussian wiretap channel with Gaussian input, shaped $Q$-ASK input and uniform $Q$-ASK input, where $Q \in \{2,4,8,16,32,64\}$ and $\SNR_B-\SNR_E=3$~dB.}
    \label{fig:Rs}
\end{figure}

Fig.~\ref{fig:Rs} depicts the achievable secrecy rate versus $\SNR_B$ for shaped \ac{ASK} constellation input with distribution $P_{X^*}$, where $P_{X^*}$ is found with a step size equal to $10^{-4}$, and $\SNR_E$ is set always to be 3 dB lower than $\SNR_B$. For comparison, the performances of Gaussian input and uniformly distributed \ac{ASK} input are also provided. It is observed in Fig.~\ref{fig:Rs} that the shaped \ac{ASK} input achieves a secrecy rate higher than the uniform case and close to the Gaussian capacity. For a certain cardinality $Q$, limited by finite constellation input, the shaped input has the same drawback as the uniform input, that is, the secrecy rate will drop to zero when $\SNR_B$ is high enough. Nevertheless, as the cardinality $Q$ of the \ac{ASK} constellation increases, the shaped \ac{ASK} input almost eliminates the gap between the Gaussian capacity and the secrecy rate achieved by the uniform \ac{ASK} input, which confirms that our solution to the optimization problem (\ref{eq:Px_opt}) is effective enough.

Note that since the \ac{QAM} signal can be generated by a pair of orthogonal \ac{ASK} signals, the method for solving optimization problem (\ref{eq:Px_opt}) in this section is also applicable to \ac{QAM} constellations. In addition, we highlight that some $P_{X^*}$ can be computed with a certain step size for \acp{SNR} beforehand and stored. Then, given an $\SNR_B$ and an $\SNR_E$, we may acquire an optimized distribution via linear fitting in practical use.


\section{Probabilistic Shaped Multilevel Polar Coding Scheme}
After optimizing the input distribution, we need a practical coding scheme to achieve the desired input distribution while ensuring secure transmission. In this section, we propose a multilevel polar coding scheme to achieve the secrecy capacity of the asymmetric \ac{DMWC} with constellation input, and also show that the scheme works for the Gaussian wiretap channel.

\subsection{Polarization of Multilevel Polar Codes for the Asymmetric DMWC}
\begin{figure}[!tp]
    \setstretch{1}
    \centering
    \tikzset{
    box1/.style ={
    rectangle, 
    minimum width = 2cm, 
    minimum height = 2cm, 
    inner sep = 6pt, 
    text centered,
    thick,
    draw = black, 
    align = center,
    },
    box2/.style ={
    rectangle, 
    minimum width = 1.5cm, 
    minimum height = 1.5cm, 
    inner sep = 6pt, 
    text centered,
    thick,
    draw = black, 
    align = center,
    },
    rec/.style ={
    rectangle, 
    minimum width = 3.4cm, 
    minimum height = 6.8cm, 
    inner sep = 6pt, 
    text centered,
    thick,
    draw = black, 
    align = center,
    },
    arrow/.style ={
    -latex,
    },
}

\begin{tikzpicture}
\node[box1] (GN1) at(0,-0.25) {$\bm{G}_N$};
\node[box1] (GN2) at([yshift=-2.3cm]GN1) {$\bm{G}_N$};
\node[box1] (GN3) at([yshift=-2.3cm]GN2) {$\bm{G}_N$};

\node[box2] (f1) at(8,0) {$f(\bm{b}_1)$};
\node[box2] (f2) at([yshift=-1.7cm]f1) {$f(\bm{b}_2)$};
\node[box2] (f3) at([yshift=-1.7cm]f2) {$f(\bm{b}_3)$};
\node[box2] (f4) at([yshift=-1.7cm]f3) {$f(\bm{b}_4)$};

\node[rec] (r) at(4.125,-2.55) {};

\draw[thick] ([yshift=0.75cm]  GN1.west) --node[xshift=-0.6cm]{$u_1^1$} ([xshift=-0.7cm,yshift=0.75cm] GN1.west);
\draw[thick] ([yshift=0.25cm]  GN1.west) --node[xshift=-0.6cm]{$u_2^1$} ([xshift=-0.7cm,yshift=0.25cm] GN1.west);
\draw[thick] ([yshift=-0.25cm] GN1.west) --node[xshift=-0.6cm]{$u_3^1$} ([xshift=-0.7cm,yshift=-0.25cm] GN1.west);
\draw[thick] ([yshift=-0.75cm] GN1.west) --node[xshift=-0.6cm]{$u_4^1$} ([xshift=-0.7cm,yshift=-0.75cm] GN1.west);

\draw[thick] ([yshift=0.75cm]  GN2.west) --node[xshift=-0.6cm]{$u_1^2$} ([xshift=-0.7cm,yshift=0.75cm] GN2.west);
\draw[thick] ([yshift=0.25cm]  GN2.west) --node[xshift=-0.6cm]{$u_2^2$} ([xshift=-0.7cm,yshift=0.25cm] GN2.west);
\draw[thick] ([yshift=-0.25cm] GN2.west) --node[xshift=-0.6cm]{$u_3^2$} ([xshift=-0.7cm,yshift=-0.25cm] GN2.west);
\draw[thick] ([yshift=-0.75cm] GN2.west) --node[xshift=-0.6cm]{$u_4^2$} ([xshift=-0.7cm,yshift=-0.75cm] GN2.west);

\draw[thick] ([yshift=0.75cm]  GN3.west) --node[xshift=-0.6cm]{$u_1^3$} ([xshift=-0.7cm,yshift=0.75cm] GN3.west);
\draw[thick] ([yshift=0.25cm]  GN3.west) --node[xshift=-0.6cm]{$u_2^3$} ([xshift=-0.7cm,yshift=0.25cm] GN3.west);
\draw[thick] ([yshift=-0.25cm] GN3.west) --node[xshift=-0.6cm]{$u_3^3$} ([xshift=-0.7cm,yshift=-0.25cm] GN3.west);
\draw[thick] ([yshift=-0.75cm] GN3.west) --node[xshift=-0.6cm]{$u_4^3$} ([xshift=-0.7cm,yshift=-0.75cm] GN3.west);

\draw[thick] ([yshift=0.75cm]  GN1.east) --node[xshift=1cm]{$b_1^1$} ([xshift=1.4cm,yshift=0.75cm] GN1.east);
\draw[thick] ([yshift=0.25cm]  GN1.east) --node[xshift=1cm]{$b_2^1$} ([xshift=1.4cm,yshift=0.25cm] GN1.east);
\draw[thick] ([yshift=-0.25cm] GN1.east) --node[xshift=1cm]{$b_3^1$} ([xshift=1.4cm,yshift=-0.25cm] GN1.east);
\draw[thick] ([yshift=-0.75cm] GN1.east) --node[xshift=1cm]{$b_4^1$} ([xshift=1.4cm,yshift=-0.75cm] GN1.east);

\draw[thick] ([yshift=0.75cm]  GN2.east) --node[xshift=1cm]{$b_1^2$} ([xshift=1.4cm,yshift=0.75cm] GN2.east);
\draw[thick] ([yshift=0.25cm]  GN2.east) --node[xshift=1cm]{$b_2^2$} ([xshift=1.4cm,yshift=0.25cm] GN2.east);
\draw[thick] ([yshift=-0.25cm] GN2.east) --node[xshift=1cm]{$b_3^2$} ([xshift=1.4cm,yshift=-0.25cm] GN2.east);
\draw[thick] ([yshift=-0.75cm] GN2.east) --node[xshift=1cm]{$b_4^2$} ([xshift=1.4cm,yshift=-0.75cm] GN2.east);

\draw[thick] ([yshift=0.75cm]  GN3.east) --node[xshift=1cm]{$b_1^3$} ([xshift=1.4cm,yshift=0.75cm] GN3.east);
\draw[thick] ([yshift=0.25cm]  GN3.east) --node[xshift=1cm]{$b_2^3$} ([xshift=1.4cm,yshift=0.25cm] GN3.east);
\draw[thick] ([yshift=-0.25cm] GN3.east) --node[xshift=1cm]{$b_3^3$} ([xshift=1.4cm,yshift=-0.25cm] GN3.east);
\draw[thick] ([yshift=-0.75cm] GN3.east) --node[xshift=1cm]{$b_4^3$} ([xshift=1.4cm,yshift=-0.75cm] GN3.east);

\draw[thick] ([yshift=0.5cm]  f1.west) --node[xshift=-1cm]{$b_1^1$} ([xshift=-1.4cm,yshift=0.5cm] f1.west);
\draw[thick] ([yshift=0.0cm]  f1.west) --node[xshift=-1cm]{$b_1^2$} ([xshift=-1.4cm,yshift=0.0cm] f1.west);
\draw[thick] ([yshift=-0.5cm] f1.west) --node[xshift=-1cm]{$b_1^3$} ([xshift=-1.4cm,yshift=-0.5cm] f1.west);

\draw[thick] ([yshift=0.5cm]  f2.west) --node[xshift=-1cm]{$b_2^1$} ([xshift=-1.4cm,yshift=0.5cm] f2.west);
\draw[thick] ([yshift=0.0cm]  f2.west) --node[xshift=-1cm]{$b_2^2$} ([xshift=-1.4cm,yshift=0.0cm] f2.west);
\draw[thick] ([yshift=-0.5cm] f2.west) --node[xshift=-1cm]{$b_2^3$} ([xshift=-1.4cm,yshift=-0.5cm] f2.west);

\draw[thick] ([yshift=0.5cm]  f3.west) --node[xshift=-1cm]{$b_3^1$} ([xshift=-1.4cm,yshift=0.5cm] f3.west);
\draw[thick] ([yshift=0.0cm]  f3.west) --node[xshift=-1cm]{$b_3^2$} ([xshift=-1.4cm,yshift=0.0cm] f3.west);
\draw[thick] ([yshift=-0.5cm] f3.west) --node[xshift=-1cm]{$b_3^3$} ([xshift=-1.4cm,yshift=-0.5cm] f3.west);

\draw[thick] ([yshift=0.5cm]  f4.west) --node[xshift=-1cm]{$b_4^1$} ([xshift=-1.4cm,yshift=0.5cm] f4.west);
\draw[thick] ([yshift=0.0cm]  f4.west) --node[xshift=-1cm]{$b_4^2$} ([xshift=-1.4cm,yshift=0.0cm] f4.west);
\draw[thick] ([yshift=-0.5cm] f4.west) --node[xshift=-1cm]{$b_4^3$} ([xshift=-1.4cm,yshift=-0.5cm] f4.west);

\draw[thick] (f1.east) --node[xshift=0.6cm]{$x_1$} ([xshift=0.7cm] f1.east);
\draw[thick] (f2.east) --node[xshift=0.6cm]{$x_2$} ([xshift=0.7cm] f2.east);
\draw[thick] (f3.east) --node[xshift=0.6cm]{$x_3$} ([xshift=0.7cm] f3.east);
\draw[thick] (f4.east) --node[xshift=0.6cm]{$x_4$} ([xshift=0.7cm] f4.east);

\draw[gray,thick] ([xshift=2cm,yshift=0.75cm]  GN1.east) -- ([xshift=-2cm,yshift=0.5cm] f1.west);
\draw[gray,thick] ([xshift=2cm,yshift=0.25cm]  GN1.east) -- ([xshift=-2cm,yshift=0.5cm] f2.west);
\draw[gray,thick] ([xshift=2cm,yshift=-0.25cm] GN1.east) -- ([xshift=-2cm,yshift=0.5cm] f3.west);
\draw[gray,thick] ([xshift=2cm,yshift=-0.75cm] GN1.east) -- ([xshift=-2cm,yshift=0.5cm] f4.west);
\draw[gray,thick] ([xshift=2cm,yshift=0.75cm]  GN2.east) -- ([xshift=-2cm,yshift=0.0cm] f1.west);
\draw[gray,thick] ([xshift=2cm,yshift=0.25cm]  GN2.east) -- ([xshift=-2cm,yshift=0.0cm] f2.west);
\draw[gray,thick] ([xshift=2cm,yshift=-0.25cm] GN2.east) -- ([xshift=-2cm,yshift=0.0cm] f3.west);
\draw[gray,thick] ([xshift=2cm,yshift=-0.75cm] GN2.east) -- ([xshift=-2cm,yshift=0.0cm] f4.west);
\draw[gray,thick] ([xshift=2cm,yshift=0.75cm]  GN3.east) -- ([xshift=-2cm,yshift=-0.5cm] f1.west);
\draw[gray,thick] ([xshift=2cm,yshift=0.25cm]  GN3.east) -- ([xshift=-2cm,yshift=-0.5cm] f2.west);
\draw[gray,thick] ([xshift=2cm,yshift=-0.25cm] GN3.east) -- ([xshift=-2cm,yshift=-0.5cm] f3.west);
\draw[gray,thick] ([xshift=2cm,yshift=-0.75cm] GN3.east) -- ([xshift=-2cm,yshift=-0.5cm] f4.west);

\end{tikzpicture}
    \caption{The multilevel polar coding construction with $Q=8$ and $N=4$.}
    \label{fig:MLPC}
\end{figure}
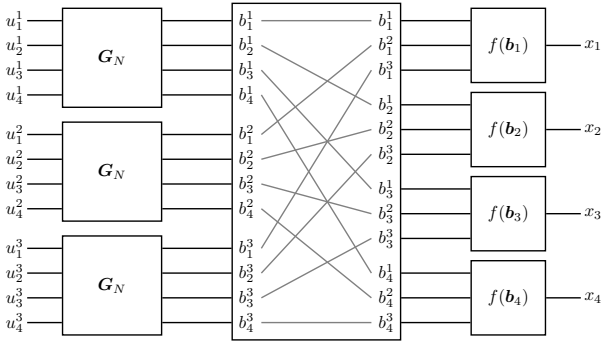

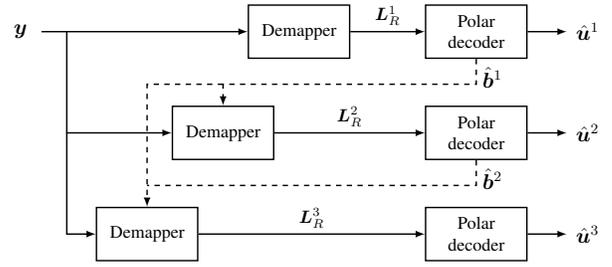
\begin{figure}[!tp]
    \setstretch{1}
    \centering
    \tikzset{
    box/.style ={
    rectangle, 
    minimum width = 2cm, 
    minimum height = 1.05cm, 
    inner sep = 4pt, 
    text centered,
    thick,
    draw = black, 
    align = center,
    },
    txt/.style ={
    rectangle, 
    minimum width = 0.8cm, 
    minimum height = 0.8cm, 
    inner sep = 0pt, 
    text centered,
    thick,
    fill = none,
    draw = none, 
    align = center,
    },
}

\begin{tikzpicture}
\node[box] (Dmap3) at(0,0) {Demapper};
\node[box] (Dmap2) at([xshift=1.5cm,yshift=2cm]Dmap3) {Demapper};
\node[box] (Dmap1) at([xshift=1.5cm,yshift=2cm]Dmap2) {Demapper};

\node[box] (Dec3) at([xshift=6.5cm]Dmap3) {Polar \\ decoder};
\node[box] (Dec2) at([xshift=5cm]Dmap2) {Polar \\ decoder};
\node[box] (Dec1) at([xshift=3.5cm]Dmap1) {Polar \\ decoder};

\node[txt] (y) at([xshift=-5.5cm]Dmap1) {\large $\bm{y}$};

\node[txt] (u1) at([xshift=2.2cm]Dec1) {\large $\hat{\bm{u}}^1$};
\node[txt] (u2) at([xshift=2.2cm]Dec2) {\large $\hat{\bm{u}}^2$};
\node[txt] (u3) at([xshift=2.2cm]Dec3) {\large $\hat{\bm{u}}^3$};

\node[txt] (b1) at([xshift=0.32cm,yshift=-0.9cm]Dec1) {\large $\hat{\bm{b}}^1$};
\node[txt] (b2) at([xshift=0.32cm,yshift=-0.9cm]Dec2) {\large $\hat{\bm{b}}^2$};

\draw[thick,-latex] (Dec1.east) -- (u1.west);
\draw[thick,-latex] (Dec2.east) -- (u2.west);
\draw[thick,-latex] (Dec3.east) -- (u3.west);

\draw[thick,-latex] (Dmap1.east) --node[above]{$\bm{L}_R^1$} (Dec1.west);
\draw[thick,-latex] (Dmap2.east) --node[above]{$\bm{L}_R^2$} (Dec2.west);
\draw[thick,-latex] (Dmap3.east) --node[above]{$\bm{L}_R^3$} (Dec3.west);

\draw[thick,-latex] (y.east) -- (Dmap1.west);
\draw[thick,-latex] ([xshift=0.5cm] y.east) -- ([xshift=0.5cm,yshift=-2cm] y.east) -- (Dmap2.west);
\draw[thick,-latex] ([xshift=0.5cm,yshift=-2cm] y.east) -- ([xshift=0.5cm,yshift=-4cm] y.east) -- (Dmap3.west);

\draw[thick,-latex,dashed] (Dec1.south) -- ([yshift=-0.5cm] Dec1.south) -- ([xshift=-5cm,yshift=-0.5cm] Dec1.south) -- (Dmap2.north);
\draw[thick,-latex,dashed] ([xshift=-5cm,yshift=-0.5cm] Dec1.south) -- ([xshift=-6.5cm,yshift=-0.5cm] Dec1.south) -- (Dmap3.north);
\draw[thick,dashed] (Dec2.south) -- ([yshift=-0.5cm] Dec2.south) -- ([xshift=-6.5cm,yshift=-0.5cm] Dec2.south);

\end{tikzpicture}
    \caption{The \ac{MSD} construction with $Q=8$.}
    \label{fig:MSD}
\end{figure}

Let $W_1 : X \to Y$ and $W_2 : X \to Z$ be Bob's channel and Eve's channel, respectively. We assume that $W_2$ is degraded with respect to $W_1$, denoted as $W_2 \preceq W_1$. The input alphabet $\mathcal{X}$ is of cardinality $Q=|\mathcal{X}|=2^q$. We consider the multilevel polar coding construction \cite{Seidl2013polar} shown in Fig.~\ref{fig:MLPC}. A codeword $\bm{x}$ consists of $N=2^n$ symbols, while each symbol $x_i$, $i \in [\![N]\!]$, is labelled with $q$ bits. We denote the symbol mapping function by $f(\cdot)$, i.e., $x_i = f(\bm{b}_i) = f(b_i^1, b_i^2, \cdots, b_i^q)$, where $f(\cdot)$ is invertible. Here, we use the set-partitioning labeling since it outperforms the classic Gary labeling in \ac{MLC} \cite{Seidl2013polar}. For each bit-level, $\bm{b}^{l}$, $l \in [\![q]\!]$, is generated by a separate polar transform $\bm{b}^{l} = \bm{u}^{l}\bm{G}_N$ of length $N$. At the receiver, the \ac{MSD} \cite{imai1977new} outputs the estimates $\hat{\bm{u}}^{l}$ of $\bm{u}^{l}$ level-by-level, as shown in Fig.~\ref{fig:MSD}. Specifically, the demapper at $l$-th bit-level calculates the \acp{LLR} $\bm{L}_R^l$ given the channel observation $\bm{y}$ and previous estimates $\hat{\bm{b}}^{[\![l-1]\!]}$, and then feeds $\bm{L}_R^l$ to a polar decoder to obtain the estimates $\hat{\bm{b}}^{l}$ and $\hat{\bm{u}}^{l}$.

With the \ac{MLC} and \ac{MSD} construction, $b^l$ is transmitted over equivalent binary-input sub-channels $W_1^l: B^l \to \left(Y, \bm{B}^{[\![l-1]\!]}\right)$ and $W_2^l: B^l \to \left(Z, \bm{B}^{[\![l-1]\!]}\right)$. The degradation of the sub-channel $W^l$ is described as follows.

\begin{lemma}
\label{Lem:DegWl}
Let $W_1 : X \to Y$ and $W_2 : X \to Z$ be two \acp{DMC} with input alphabet $\mathcal{X}$ of cardinality $|\mathcal{X}|=2^q$. The input symbol $X$ is mapped by $f(\bm{B}) = f(B^1, B^2, \cdots, B^q)$. Consider the equivalent binary-input sub-channels $W_1^l: B^l \to \left(Y, \bm{B}^{[\![l-1]\!]}\right)$ and $W_2^l: B^l \to\left(Z, \bm{B}^{[\![l-1]\!]}\right)$ for the $l$-th bit-level, $l \in [\![q]\!]$. If $W_2 \preceq W_1$, then $W_2^l \preceq W_1^l$.
\end{lemma}
\begin{IEEEproof}
    See Appendix \ref{Prof:Lem:DegWl}.
\end{IEEEproof}

Hence, $b^l$ is transmitted over a degraded binary-input \ac{DMWC}. By using polar transform $\bm{b}^{l} = \bm{u}^{l}\bm{G}_N$, the degradation of the $i$-th bit-channel $W^{l(i)}_N$ at the $l$-th bit-level is given by the following lemma. 

\begin{lemma}[Degradation of $\bar{W}_N^{(i)}$ {\cite[Lemma 4.7]{Korada2009polar}}]
\label{Lem:DegW}
Let $\bar{W}_1 : X \to Y$ and $\bar{W}_2 : X \to Z$ be two \acp{B-DMC} such that $\bar{W}_2 \preceq \bar{W}_1$, then for all $i \in [\![N]\!]$, $\bar{W}_{2,N}^{(i)} \preceq \bar{W}_{1,N}^{(i)}$. 
\end{lemma}

From Lemma \ref{Lem:DegW}, we immediately have that $W^{l(i)}_{2,N} \preceq W^{l(i)}_{1,N}$ for $l \in [\![q]\!]$ and $i \in [\![N]\!]$. Then, in the following lemma, we show that the degradation of the bit-channel $W^{l(i)}_{N}$ will increase the Bhattacharyya parameter.

\begin{lemma}
\label{Lem:ZB}
Let $\bar{W}_1 : X \to Y$ and $\bar{W}_2 : X \to Z$ be two \acp{B-DMC}. Consider a polar transform $\bm{X} = \bm{U}\bm{G}_N$. Let $\bar{W}_{1,N}^{(i)} : U_i \to \bm{Y}$ and $\bar{W}_{2,N}^{(i)}: U_i \to \bm{Z}$ be two bit-channels with side information $\bm{S}$ for $i \in [\![N]\!]$. If $\bar{W}_{2,N}^{(i)} \preceq \bar{W}_{1,N}^{(i)}$, we then have
$$Z_B\left(U_i | \bm{S}, \bm{Y} \right) \leq Z_B\left(U_i | \bm{S}, \bm{Z} \right).$$
\end{lemma}
\begin{IEEEproof}
    See Appendix \ref{Prof:Lem:ZB}.
\end{IEEEproof}

Now, we consider the sets
\begin{align}
    \label{eq:LX}
    \mathcal{L}_X &= \left\{(l,i): Z_B\left(U_i^l| \bm{V}_i^l\right) < \delta_N \right\}, \\
    \label{eq:HX}
    \mathcal{H}_X &= \left\{(l,i): Z_B\left(U_i^l| \bm{V}_i^l\right) > 1-\delta_N \right\}, \\
    \label{eq:LXY}
    \mathcal{L}_{X|Y} &= \left\{(l,i): Z_B\left(U_i^l| \bm{V}_i^l, \bm{Y}\right) < \delta_N \right\}, \\
    \label{eq:HXY}
    \mathcal{H}_{X|Y} &= \left\{(l,i): Z_B\left(U_i^l| \bm{V}_i^l, \bm{Y}\right) > 1-\delta_N \right\}
\end{align}    
where $l \in [\![q]\!]$, $i \in [\![N]\!]$, and $\bm{V}_i^l = (\bm{U}^l_{[\![i-1]\!]}, \bm{B}^{[\![l-1]\!]}_{[\![N]\!]} )$. Recall the fact that $Z_B(\cdot|\cdot)$ and $H(\cdot|\cdot)$ are close to 0 or 1 simultaneously. Therefore, $U_i^l$ is approximately a deterministic function of $\bm{V}_i^l$ and $\bm{Y}$ for $(l,i) \in \mathcal{L}_{X|Y}$, while $U_i^l$ is approximately uniformly distributed and independent of $\bm{V}_i^l$ and $\bm{Y}$ for $(l,i) \in \mathcal{H}_{X|Y}$. In other words, the bit positions indexed by $\mathcal{L}_{X|Y}$ are suitable for data transmission, since we can successively recover $U_i^l$ given the channel output $\bm{Y}$ and the sequence $\bm{V}_i^l$ with \ac{MSD} and \ac{SC} decoding. Meanwhile, for $(l,i) \in \mathcal{H}_{X}$, $U_i^l$ is almost independent of $\bm{V}_i^l$, while $U_i^l$ is almost a deterministic function of $\bm{V}_i^l$ for $(l,i) \in \mathcal{L}_{X}$. Hence, if $(l,i) \in \mathcal{L}_{X}$, $U_i^l$ can not be used for data transmission, but needs to be successively chosen in a specific way based on the previous input $\bm{V}_i^l$. The polarization of these sets is given as follows.

\begin{proposition}
\label{Prop:HXY}
Let $W : X \to Y$ be a \ac{DMC} with input alphabet $\mathcal{X}$ of cardinality $|\mathcal{X}|=2^q$. Let $B^l$, $l \in [\![q]\!]$, be the $l$-th bit of the binary representation $\bm{B}$ of symbol $X = f(\bm{B})$, and let $\bm{B}^l = \bm{U}^l \bm{G}_N$. The sets $\mathcal{L}_X$, $\mathcal{H}_X$, $\mathcal{L}_{X|Y}$ and $\mathcal{H}_{X|Y}$ are defined by (\ref{eq:LX})-(\ref{eq:HXY}), respectively. As $N \to\infty$, we asymptotically have
\begin{align}
    \label{eq:NLHX}
    q - \lim_{N \to \infty} \frac{1}{N} \big| \mathcal{L}_{X} \big| &= \lim_{N \to \infty} \frac{1}{N} \big| \mathcal{H}_{X} \big| \quad = H(X), \\
    \label{eq:NLHXY}
    q - \lim_{N \to \infty} \frac{1}{N} \big| \mathcal{L}_{X|Y} \big| &= \lim_{N \to \infty} \frac{1}{N} \big| \mathcal{H}_{X|Y} \big| = H(X|Y).
\end{align}
\end{proposition}
\begin{IEEEproof}
    See Appendix \ref{Prof:Prop:HXY}.
\end{IEEEproof}

Let $\mathcal{L}_{X|Z}$ and $\mathcal{H}_{X|Z}$ be the two sets defined in the similar way as $\mathcal{L}_{X|Y}$ and $\mathcal{H}_{X|Y}$, respectively. By Lemma \ref{Lem:ZB} and the fact that conditioning reduces the Bhattacharyya parameter \cite[Lem. 6]{liu2018construction}, we have
\begin{equation}
    \mathcal{L}_{X} \subseteq \mathcal{L}_{X|Z} \subseteq \mathcal{L}_{X|Y}.
    \label{eq:LXLXZLXY}
\end{equation}
This reveals that although the bits at positions indexed by $\mathcal{L}_{X|Y}$ can be successively recovered by Bob, the bits at positions indexed by $\mathcal{L}_{X|Z}$ among them will be directly leaked to Eve. To achieve secure transmission, not all bit positions indexed by $\mathcal{L}_{X|Y}$ are suitable for message transmission, where the bit positions indexed by $\mathcal{L}_{X|Z}$ can only be filled with some useless information to Eve.

Thus, we divide the index set $\{(l,i): l\in [\![q]\!], i \in [\![N]\!]\}$ of a multilevel polar code into four sets:
\begin{align}
    \label{eq:F}
    \mathcal{F} &= \mathcal{L}_{X|Y}^{C}, \\
    \label{eq:A}
    \mathcal{A} &= (\mathcal{L}_{X|Y} \cap \mathcal{H}_X) \setminus (\mathcal{L}_{X|Z} \cap \mathcal{H}_X), \\
    \label{eq:R}
    \mathcal{R} &= \mathcal{L}_{X|Z} \cap \mathcal{H}_{X}, \\ 
    \label{eq:D}
    \mathcal{D} &= \mathcal{L}_{X|Y} \cap \mathcal{H}_{X}^{C}. 
\end{align}
It is immediately obtained that the sets $\mathcal{F}$, $\mathcal{A}$, $\mathcal{R}$, $\mathcal{D}$ are disjoint and $\mathcal{F} \cup \mathcal{A} \cup \mathcal{R} \cup \mathcal{D} = \{(l,i): l\in [\![q]\!], i \in [\![N]\!]\}$ by (\ref{eq:LXLXZLXY}). The polarization of such a polar code is described as follows.
\begin{theorem}
\label{Thm:ML-FARD}
Let $W_1 : X \to Y$ and $W_2 : X \to Z$ be two \acp{DMC} such that $W_2 \preceq W_1$. The input alphabet $\mathcal{X}$ is of cardinality $|\mathcal{X}|=2^q$. Let $B^l$, $l \in [\![q]\!]$, be the $l$-th bit of the binary representation $\bm{B}$ of symbol $X = f(\bm{B})$, and let $\bm{B}^l = \bm{U}^l \bm{G}_N$. Given the sets $\mathcal{F}$, $\mathcal{A}$, $\mathcal{R}$ and $\mathcal{D}$ defined by (\ref{eq:F})-(\ref{eq:D}), respectively, we asymptotically have\vspace{0cm}
\begin{align}
    \lim_{N \to \infty} \frac{1}{N} \left|\mathcal{F}\right| &= H(X|Y), \\
    \lim_{N \to \infty} \frac{1}{N} \left|\mathcal{A}\right| &= I(X;Y) - I(X;Z), \\
    \label{eq:NR_IXY}
    \lim_{N \to \infty} \frac{1}{N} \left|\mathcal{R}\right| &= I(X;Z), \\
    \lim_{N \to \infty} \frac{1}{N} \left|\mathcal{D}\right| &= q-H(X).
\end{align}
\end{theorem}
\begin{IEEEproof}
    See Appendix \ref{Prof:Thm:ML-FARD}.
\end{IEEEproof}

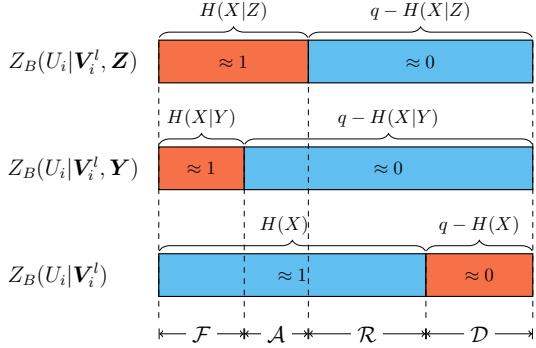
\begin{figure}[!t]
    \centering
    \tikzset{
    box/.style ={
    rectangle, 
    minimum width = 3cm, 
    minimum height = 1cm, 
    inner sep = 5pt, 
    draw = none, 
    align = center,
    },
    brace/.style ={
    decorate,
    decoration={brace,amplitude=2mm},
    },
}

\begin{tikzpicture}
\draw[fill=Red, line width = 0.6pt] (0,0) rectangle node(Z1){$\approx 1$} (2.8,-0.8);
\draw[fill=Grn, line width = 0.6pt] (2.8,0) rectangle node(Z0){$\approx 0$} (7,-0.8);

\draw[fill=Red, line width = 0.6pt] [yshift=-2cm] (0,0) rectangle node(Y1){$\approx 1$} (1.6,-0.8);
\draw[fill=Grn, line width = 0.6pt] [yshift=-2cm] (1.6,0) rectangle node(Y0){$\approx 0$} (7,-0.8);

\draw[fill=Grn, line width = 0.6pt] [yshift=-4cm] (0,0) rectangle node(X1){$\approx 1$} (5,-0.8);
\draw[fill=Red, line width = 0.6pt] [yshift=-4cm] (5,0) rectangle node(X0){$\approx 0$} (7,-0.8);

\node[box] (HXZ) at ([xshift=-3cm]Z1)  {\large{$Z_B(U_i|\bm{V}_i^l,\bm{Z})$}};
\node[box] (HXY) at ([yshift=-2cm]HXZ) {\large{$Z_B(U_i|\bm{V}_i^l,\bm{Y})$}};
\node[box] (HX)  at ([yshift=-2cm]HXY) {\large{$Z_B(U_i|\bm{V}_i^l)$\quad~}};

\draw[brace] (0.01,0.05) -- (2.79,0.05);
\draw[brace] (2.81,0.05) -- (6.99,0.05);
\node[box] at ([yshift=0.7cm] Z1.north) {$H(X|Z)$};
\node[box] at ([yshift=0.7cm] Z0.north) {$q-H(X|Z)$};

\draw[brace] [yshift=-2cm] (0.01,0.05) -- (1.59,0.05);
\draw[brace] [yshift=-2cm] (1.61,0.05) -- (6.99,0.05);
\node[box] at ([yshift=0.7cm] Y1.north) {$H(X|Y)$};
\node[box] at ([yshift=0.7cm] Y0.north) {$q-H(X|Y)$};

\draw[brace] [yshift=-4cm] (0.01,0.05) -- (4.99,0.05);
\draw[brace] [yshift=-4cm] (5.01,0.05) -- (6.99,0.05);
\node[box] at ([xshift=-0.15cm,yshift=0.7cm] X1.north) {$H(X)$};
\node[box] at ([yshift=0.7cm] X0.north) {$q-H(X)$};

\draw [dashed] (0,-5.2) -- (0,0);
\draw [dashed] (1.6,-5.2) -- (1.6,-2);
\draw [dashed] (2.8,-5.2) -- (2.8,0);
\draw [dashed] (5,-5.2) -- (5,-4);
\draw [dashed] (7,-5.2) -- (7,0);

\draw [|<->|] (0,-5.5) --node[fill=white]{\large$\mathcal{F}$} (1.6,-5.5);
\draw [|<->|] (1.6,-5.5) --node[fill=white]{\large$\mathcal{A}$} (2.8,-5.5);
\draw [|<->|] (2.8,-5.5) --node[fill=white]{\large$\mathcal{R}$} (5,-5.5);
\draw [|<->|] (5,-5.5) --node[fill=white]{\large$\mathcal{D}$} (7,-5.5);

\end{tikzpicture}
    \caption{Graphical representation of the polarization of $Z_B(U_i|\bm{V}_i^l,\bm{Z})$, $Z_B(U_i|\bm{V}_i^l,\bm{Y})$ and $Z_B(U_i|\bm{V}_i^l)$ as $N \to \infty$.}
    \label{fig:polar}
\end{figure}

A graphical representation of the polarization is illustrated in Fig.~\ref{fig:polar}. If the bit positions $\mathcal{A}$ are placed with uniformly distributed message, Theorem \ref{Thm:ML-FARD} and (\ref{eq:Rs}) tell us that the multilevel polar codes above can asymptotically achieve the secrecy rate $R_s$ when the input distribution $P_X$ is given. Note that if $P_X$ is an optimal solution to (\ref{eq:Px_opt}), such polar codes can be proved to achieve the secrecy capacity of asymmetric \ac{DMWC} with constellation input by Theorem \ref{Thm:ML-FARD} and (\ref{eq:Cs}). However, our optimized sub-optimal input distribution $P_{X^*}$ will remain a negligible gap between the achievable secrecy rate and the security capacity as verified in Sec. \ref{Sec:PxOpt}.

\subsection{Proposed Coding Scheme}
Based on the polarization phenomenon above, the general idea of our coding scheme for asymmetric \ac{DMWC} with constellation input is summarized as follows.
\begin{itemize}
    \item The positions indexed by $\mathcal{A}$ are used for transmitting uniformly distributed message $\bm{M}$.
    \item The positions indexed by $\mathcal{R}$ are placed with uniform \textit{random bits}. The random bits here are to keep Eve ignorant since Eve has the ability to successfully recover these random bits.
    \item The positions indexed by $\mathcal{D}$ are placed with \textit{shaping bits}, which are chosen successively according to the \textit{randomized rounding rule} \cite{honda2013polar} or the \textit{argmax rule} \cite{mondelli2018achieve}. We will describe this in more detail in Section \ref{Sec:EncDec}. 
    \item The remaining positions, i.e., indexed by $\mathcal{F}$, are filled with \textit{frozen bits}, which are set to fixed values that are known both at the transmitter and the receiver. 
\end{itemize}

\subsection{Reliability Analysis}
Recall that $\bm{M}$ and $\bm{M}_B$ denote Alice's message and Bob's observation, respectively. Our coding scheme satisfies the reliability condition (\ref{eq:r_condt}) by the following theorem.
\begin{theorem}
\label{Thm:Relb}
Let $W_1 : X \to Y$ and $W_2 : X \to Z$ be two \acp{DMC} with constellation input such that $W_2 \preceq W_1$. Consider the coding scheme proposed above. For any $0<\beta^\prime<\beta<1/2$, the average block error probability is upper bounded by
\begin{equation}
\label{eq:PeBd}
    \mathrm{Pr}\{\hat{\bm{M}}_B \neq \bm{M} \} \leq c2^{-N^{\beta^\prime}}
\end{equation}
where $c$ is a positive constant. As $K \to \infty$, we have
\begin{equation}
\label{eq:Relb}
    \lim_{K\to \infty} \mathrm{Pr}\{\hat{\bm{M}}_B \neq \bm{M} \} = 0.
\end{equation}
\end{theorem}
\begin{IEEEproof}
Since $\mathcal{A} \subseteq (\mathcal{L}_{X|Y} \cap \mathcal{H}_X)$, there is a positive constant $c$ so that (\ref{eq:PeBd}) holds with $0<\beta^\prime<\beta<1/2$ by \cite[Theorem 2]{runge2022multilevel}. Then, since $K = \varTheta(N)$, it is easy to prove that (\ref{eq:Relb}) holds.
\end{IEEEproof}

\subsection{Security Analysis}
Observing that $\bm{U}_{\mathcal{A}}=\bm{M}$, $\bm{U}_{\mathcal{F}}=\bm{0}$ (for simplicity), $\bm{U}_{\mathcal{R}}$ is uniform over $\{0,1\}^{|\mathcal{R}|}$, and $\bm{U}_{\mathcal{D}}$ is determined by $\bm{U}_{\mathcal{D}^C}$. Let $h_2(\cdot)$ be the binary entropy function. We first present the following lemma that is helpful to the proof of security.
\begin{lemma}
\label{Lem:HRZA}
\begin{equation}
    H(\bm{U}_\mathcal{R}|\bm{Z},\bm{U}_\mathcal{A}) \leq h_2(c2^{-N^{\beta^\prime}}) + c|\mathcal{R}|2^{-N^{\beta^\prime}}.
\end{equation}
\end{lemma}
\begin{IEEEproof}
    See Appendix \ref{Prof:Lem:HRZA}.
\end{IEEEproof}

Now, we show that our coding scheme satisfies the weak security condition (\ref{eq:ws_condt}) in the following theorem.
\begin{theorem}
\label{Thm:Secr}
Let $W_1 : X \to Y$ and $W_2 : X \to Z$ be two \acp{DMC} with constellation input such that $W_2 \preceq W_1$. The information leakage per symbol of the proposed coding scheme vanishes as $N\to\infty$, namely 
\begin{equation}
    \lim_{N\to\infty} \frac{I(\bm{M};\bm{Z})}{N} = 0.
\end{equation}
\end{theorem}
\begin{IEEEproof}
    See Appendix \ref{Prof:Thm:Secr}.
\end{IEEEproof}

\begin{remark}
Along the lines of \cite[Part \uppercase\expandafter{\romannumeral 4}, Appendix 7]{shannon1948mathematical}, Theorems \ref{Thm:ML-FARD}, \ref{Thm:Relb} and \ref{Thm:Secr} can be easily extended to discrete-input, continuous-output degraded wiretap channels, such as the degraded Gaussian wiretap channel with constellation input.
\end{remark}


\section{Coding for Finite Block Length}
At finite block length regimes, the coding procedure and a general code construction method are introduced as follows.

\subsection{Encoding and Decoding} \label{Sec:EncDec}
At the encoder, the bit positions $\mathcal{A}$, $\mathcal{R}$ and $\mathcal{F}$ in $\bm{u}$ are directly placed with information bits, uniform random bits and frozen bits, respectively, while $u^l_i$, $(l,i) \in \mathcal{D}$, is chosen by performing an \ac{MSD} with \ac{SC} decoding. More specifically, for each bit-level $l$, the \ac{MSD} computes $P_{B^l|\bm{B}^{[\![l-1]\!]}} (\cdot|\bm{b}^{[\![l-1]\!]}_i)$ given a desired input distribution $P_X$, $i \in [\![N]\!]$ \cite{imai1977new}. Then, the \ac{SC} decoder is initialized with a priori \ac{LLR}
\begin{equation}
    L_R = \log \frac{P_{B^l|\bm{B}^{[\![l-1]\!]}} (0|\bm{b}^{[\![l-1]\!]}_i)}{P_{B^l|\bm{B}^{[\![l-1]\!]}} (1|\bm{b}^{[\![l-1]\!]}_i)}
\end{equation}
for all $i \in [\![N]\!]$, and successively outputs the probability $P_{U^l_i|\bm{V}^l_i} (\cdot|\bm{v}^l_i)$. Honda and Yamamoto proposed to determine $u^l_i$, $(l,i) \in \mathcal{D}$, by the following randomized rounding rule \cite{honda2013polar}:
\begin{equation}
    u^l_i=\left\{\begin{aligned}
    0 \quad \text{with probability}~P_{U^l_i|\bm{V}^l_i} \left(0 | \bm{v}^l_i\right), \\
    1 \quad \text{with probability}~P_{U^l_i|\bm{V}^l_i} \left(1 | \bm{v}^l_i\right)~
    \end{aligned}\right.
    \label{eq:randomized_rule}
\end{equation}
where the random number generator is shared between the transmitter and the receiver. 
To avoid randomness, the authors in \cite{mondelli2018achieve} introduced an alternative rule called argmax rule, which is given by
\begin{equation}
    u^l_i=\left\{\begin{aligned}
    0, \quad P_{U^l_i|\bm{V}^l_i} \left(0 | \bm{v}^l_i\right) \geq P_{U^l_i|\bm{V}^l_i} \left(1 | \bm{v}^l_i\right), \\
    1, \quad P_{U^l_i|\bm{V}^l_i} \left(0 | \bm{v}^l_i\right) < P_{U^l_i|\bm{V}^l_i} \left(1 | \bm{v}^l_i\right).
    \end{aligned}\right.
    \label{eq:argmax_rule}
\end{equation}
After all the values in $\bm{u}$ are determined, the codeword $\bm{x}$ is generated by the polar transform $\bm{b}^{l} = \bm{u}^{l}\bm{G}_N$ and the symbol mapping function $f(\bm{b})$.

However, the hard decision of $u^l_i$ may not achieve the desired distribution of $X$ at finite block length regimes. To alleviate this issue, we 
perform a list encoding by replacing the \ac{SC} decoder with a \ac{SCL} decoder. As such, we obtain a list of valid codewords and choose the one that has an empirical distribution closest to the desired distribution as the output of the list encoder.

\begin{figure*}[t]
    \centering
    \captionsetup[subfloat]{justification=centering}
    \subfloat[8-ASK, $K=400$, $\SNR_B=13$ dB, \\ and $d\SNR_g=4.27$ dB.]{
        \includegraphics[width=0.32\textwidth]{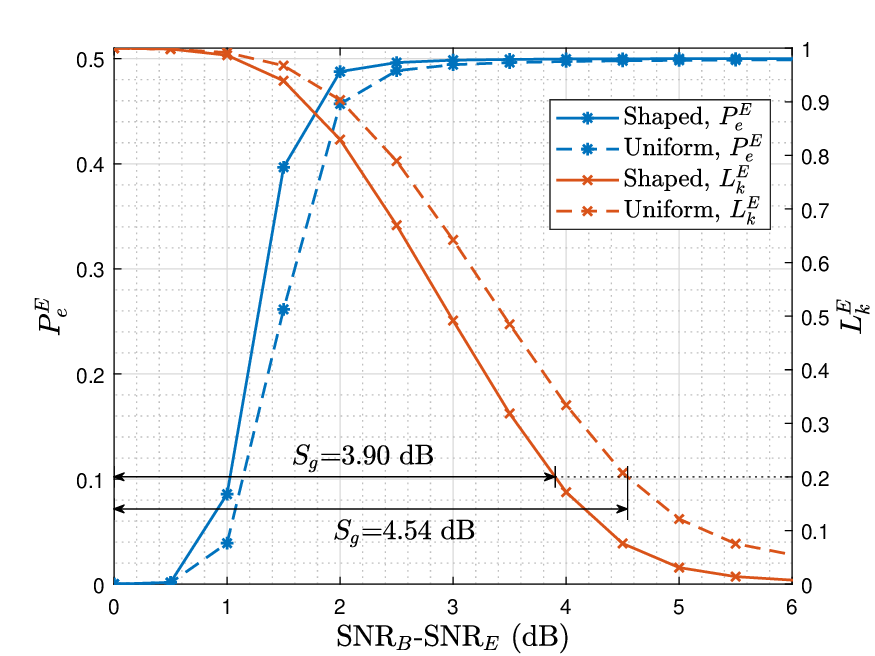}
    }\hspace{-0.2cm}
    \subfloat[16-ASK, $K=600$, $\SNR_B=18$ dB, \\ and $d\SNR_g=5.4$ dB.]{
        \includegraphics[width=0.32\textwidth]{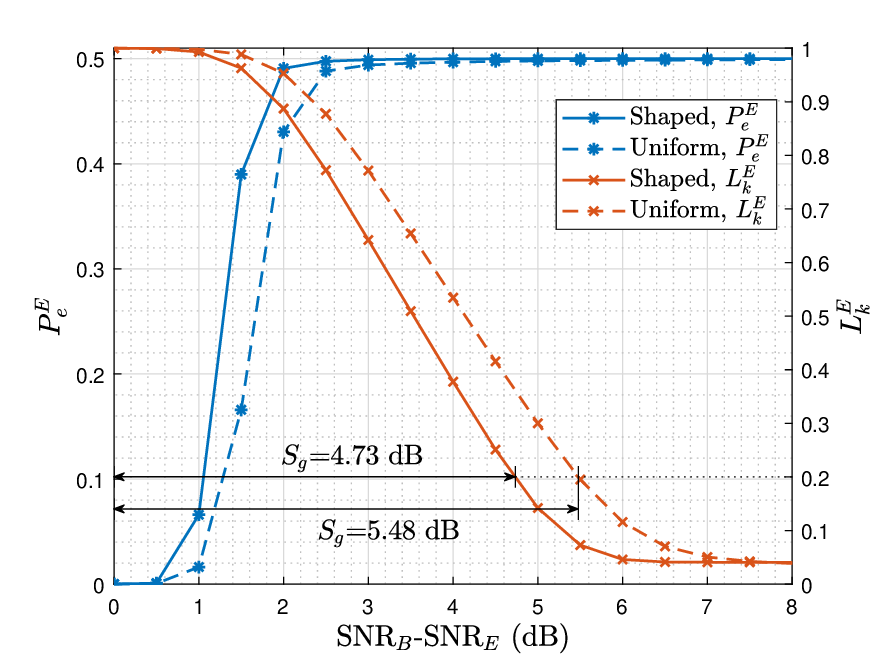}
    }\hspace{-0.2cm}
    \subfloat[64-QAM, $K=400$, $\SNR_B=13$ dB, \\ and $d\SNR_g=2.9$ dB.]{
        \includegraphics[width=0.32\textwidth]{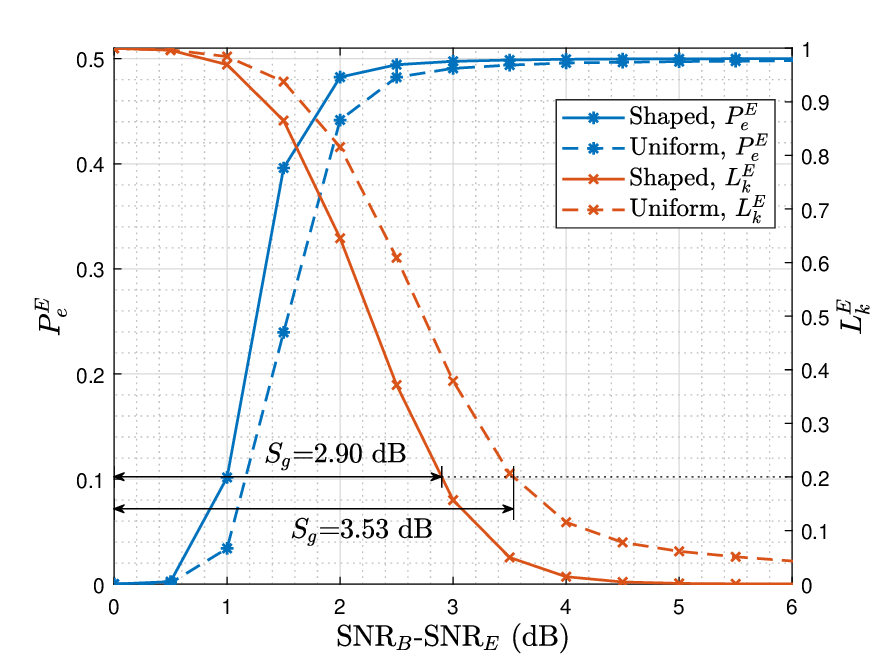}
    }
    \caption{Eve's BER $P_e^E$ and the estimated normalized information leakage $L_k^E$ versus $\SNR_B-\SNR_E$ for different modulation and message length $K$ over the Gaussian wiretap channel, where $L_{k,\max}^E = 0.2$, $P_{e,\max}^B = 10^{-5}$, and $N = 1024$.}
    \label{fig:simu_BER}
\end{figure*}

As Bob and Eve perform the same decoding procedure, we take Bob's decoder as an example. The decoder computes the estimate $\hat{\bm{u}}$ of $\bm{u}$ from the channel observations $\bm{y}$ by using the \ac{MSD} structure with \ac{SC} decoding or \ac{SCL} decoding. For $(l,i) \in \mathcal{F}$, $\hat{u}^l_i$ is set to the frozen value. Otherwise, $\hat{u}^l_i$ is estimated according to 
\begin{equation}
\label{eq:arg_u}
    \hat{u}^l_i = \arg\max_{u\in\{0,1\}} P_{U^l_i|\bm{V}^l_i,\bm{Y}} \left(u | \bm{v}^l_i, \bm{y}\right).
\end{equation}
Note that since the random bits with index set $\mathcal{R}$ are unknown to the receiver, we treat them as message bits when decoding. Moreover, $\hat{u}^l_i$ with $(l,i) \in \mathcal{D}$ should actually be estimated by $\arg\max_{u\in\{0,1\}} P_{U^l_i|\bm{V}^l_i,\bm{Y}} \left(u | \bm{v}^l_i\right)$\vspace{0.05cm}. Yet, by the fact that $H\left(U^l_i | \bm{V}^l_i, \bm{Y}\right) \leq H\left(U^l_i | \bm{V}^l_i\right)$\vspace{0.05cm}, it holds that $H\left(U^l_i | \bm{V}^l_i, \bm{Y}\right)$ is close to zero if $H\left(U^l_i | \bm{V}^l_i\right)$ is close to zero. Hence, we can still estimate $\hat{u}^l_i$ reliably by (\ref{eq:arg_u}) instead of running an additional decoder.
Furthermore, when applying \ac{SCL} decoding in \ac{MSD}, the list and path metric of the \ac{SCL} decoder at each bit-level should to be passed to the next level. To address this, the \ac{MSD} with \ac{SCL} decoding performs the same steps as the list multistage decoder in \cite{icscan2020sign}.

\subsection{Code Construction} \label{Sec:CCon}
For convenience, we first denote $|\mathcal{F}|$, $|\mathcal{A}|$, $|\mathcal{R}|$ and $|\mathcal{D}|$ by $F$, $K$, $R$, and $D$, respectively. The code construction includes the determination of the values of $F$, $K$, $R$, and $D$ and the selection of the sets $\mathcal{F}$, $\mathcal{A}$, $\mathcal{R}$, and $\mathcal{D}$.

Let $d\SNR_B$ and $d\SNR_g$ be the desired \acp{SNR} that Bob's BER $P_e^B$ at $d\SNR_B$ satisfies the reliability condition $P_e^B \leq P_{e,\max}^B$ and information leakage $L_k^E$ at $d\SNR_B-d\SNR_g$ satisfies the security condition $L_k^E \leq L_{k,\max}^E$, where $P_{e,\max}^B$ and $L_{k,\max}^E$ are the thresholds introduced in Sec. \ref{Sec:PreSm}. Observing that $P_e^B$ is an increasing function of $K+R$ as $u^l_i$, $(l,i) \in \mathcal{A}\cup\mathcal{R}$, is unknown to Bob, we have a maximum value of $K+R$ that satisfies the reliability condition. Then, $R$ should be determined under the security condition constraints. Since imposing high \ac{BER} at Eve does not guarantee any information theoretic security, we adopt the leakage-based security condition instead of the \ac{BER}-based security condition suggested in \cite{klinc2011ldpc}. The information leakage can be estimated by the Tal-Vardy construction with the upgrading approximation and the parameter $\mu = 64$ for polar codes, which offers an upper bound on the capacity of polarized bit-channels \cite{tal2013construct}. Thus, the normalized information leakage is calculated by
\begin{equation}
    L_k^E = \frac{1}{K} \sum_{(l, i) \in\mathcal{A}} C\left(W^{l(i)}_{2,N}\right)
\end{equation}
where $C(W^{l(i)}_{2,N})$ is the estimated capacity of the $i$-th polarized bit-channel at the $l$-th bit-level of Eve's channel. Note that as $R$ increases, $L_k^E$ will decrease due to the selection method of $\mathcal{R}$ in the following paragraph. We choose $R$ to be the minimum value such that the security condition is satisfied for efficient and secure transmission. However, if the value of $K$ is forcibly given, we may need to sacrifice some security to obtain an $R$. Now, the desired input distribution is set as the optimized input distribution $P_{X^*}$ resulting in a mutual information $I(X;Y)$ equal to $(K+R)/N$ for fixed $\SNR_B-\SNR_E=d\SNR_g$ across $\SNR_B$. Thus, we choose the value of $D$ to be $\kappa_D (q-H(X))$, where $\kappa_D$ is a scaling factor for better shaping performance \cite{wiegart2019shaped}. Finally, $F$ is calculated as $F=qN-K-R-D$.

The index sets $\mathcal{F}$, $\mathcal{A}$, $\mathcal{R}$, and $\mathcal{D}$ are selected according to the sort of $H\left(U^l_i | \bm{V}^l_i\right)$, $H\left(U^l_i | \bm{V}^l_i, \bm{Y}\right)$ and $H\left(U^l_i | \bm{V}^l_i, \bm{Z}\right)$. We can use the Monte-Carlo approach to estimate these entropies. To estimate $H\left(U^l_i | \bm{V}^l_i\right)$, we perform the \ac{MSD} with \ac{SC} decoding similar to that used in the encoder, that is, the \ac{SC} decoder at each bit-level is initialized with a priori \acp{LLR}. By choosing $u^l_i$ successively with the randomized rounding rule (\ref{eq:randomized_rule}) for all $l\in[\![q]\!]$ and $i\in[\![N]\!]$, the decoder successively outputs $P_{U^l_i|\bm{V}^l_i} (\cdot|\bm{v}^l_i)$. After many trials, we obtain the estimates of $H\left(U^l_i | \bm{V}^l_i\right)$. As for the estimates of $H\left(U^l_i | \bm{V}^l_i, \bm{Y}\right)$, we repeat the transmission over the channel, where the transmitter sends randomly generated data and the receivers perform an \ac{MSD} with \ac{SC} decoding. When the \ac{SC} decoder makes a wrong decision on $u^l_i$, we count the number of errors at this bit position and correct the error before moving to the next bit position. After sampling over enough frames, we calculate the error rate $\varepsilon^l_i$ for each bit position. Since $H\left(U^l_i | \bm{V}^l_i, \bm{Y}\right)$ is a monotonically increasing function of $\varepsilon^l_i$, the sort of $H\left(U^l_i | \bm{V}^l_i, \bm{Y}\right)$ is equivalent to the sort of $\varepsilon^l_i$ for all bit positions. Finally, the estimate of $H\left(U^l_i | \bm{V}^l_i, \bm{Z}\right)$ is the same as that of $H\left(U^l_i | \bm{V}^l_i, \bm{Y}\right)$. 

Now, the $F$ bit positions with the highest $H\left(U^l_i | \bm{V}^l_i, \bm{Y}\right)$ are selected into the index set $\mathcal{F}$ and the $D$ bit positions with the lowest $H\left(U^l_i | \bm{V}^l_i\right)$ are selected into the index set $\mathcal{D}$. Among the remaining positions, the $R$ bit positions with the lowest $H\left(U^l_i | \bm{V}^l_i, \bm{Z}\right)$ form the index set $\mathcal{R}$ and the remains form the index set $\mathcal{A}$.

We also remark that if the polar codes are constructed for \ac{AWGN} channel, one can follow the idea of binary-input \ac{AWGN} surrogates proposed in \cite{wu2023conditional} and perform the Tal-Vardy construction \cite{tal2013construct} or Gaussian approximation construction \cite{dai2017does} on each bit-level to estimate the entropies with less amount of computation. Moreover, if we construct for \ac{RF} channel, the \ac{RF} channel can be equivalent to an AWGN channel with the same ergodic average achievable rate as the \ac{RF} channel \cite{zhou2019construction}. Thus, the polar construction for \ac{RF} channel is transformed to the construction for AWGN channel.

\subsection{Complexity Analysis}
The encoding and decoding procedures of the proposed coding scheme leverage conventional \ac{MSD} and \ac{SCL} decoding, without the need for any additional operations. At the $l$-th bit-level, the encoder first traverses $2^{q-l+1}$ channel inputs $x$ for each possible bit label $\bm{b}^{[\![l-1]\!]}$ to calculate the probability $P_{B^l|\bm{B}^{[\![l-1]\!]}} (\cdot|\bm{b}^{[\![l-1]\!]})$, followed by an \ac{SCL} decoder with list size $L_e$. Thus, the overall encoding complexity can be represented as $\mathcal{O}\left(q(Q + L_eN\log_2(N))\right)$. As for decoding, a \ac{MAP} demapper, which needs to traverses $2^{q-l+1}$ channel inputs $x$ for $N$ channel observations $y$ at the $l$-th bit-level, and an \ac{SCL} decoder with list size $L_d$ are performed. The resulting decoding complexity is $\mathcal{O}\left(qN(2^{q-l+1} + L_d\log_2(N))\right)$.

\section{Simulation Results}
\begin{figure}[!t]
    \centering
    \includegraphics[width=0.42\textwidth]{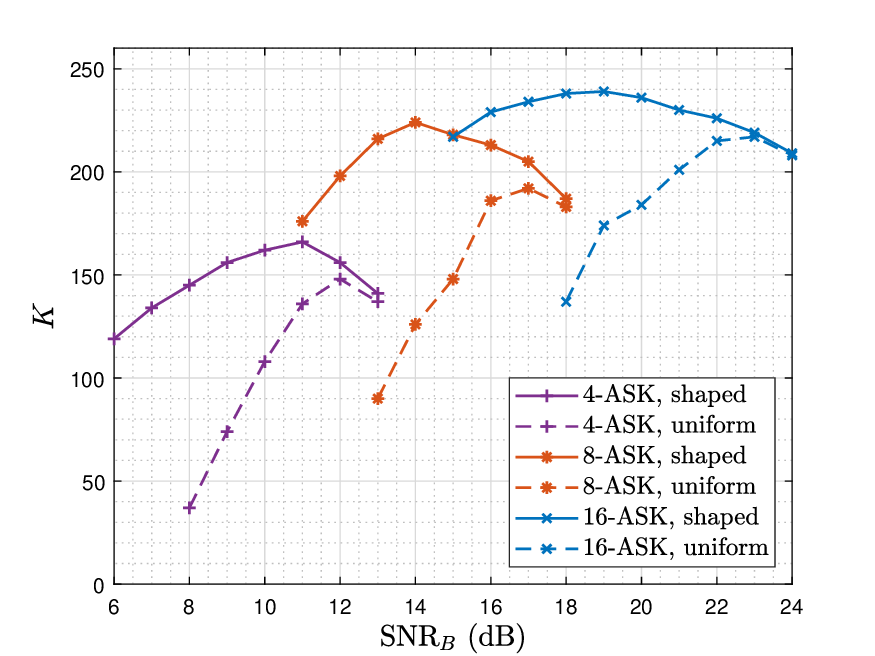}
    \caption{The maximum achievable message length $K$ for $Q$-\ac{ASK} input over the Gaussian wiretap channel under constraints of $S_g=3$ dB, $P_{e,\max}^B = 10^{-5}$, and $L_{k,\max}^E = 0.2$, where $Q \in \{4,8,16\}$ and $N=1024$.}
    \label{fig:K3dB}
\end{figure}
In this section, we evaluate the performance of the proposed scheme with Monte-Carlo simulations. 
In the following simulations, we perform the Tal-Vardy construction \cite{tal2013construct} with the degrading approximation and the parameter $\mu = 32$ for polar construction. Let $\SNR_B^*$ and $\SNR_E^*$ be the \ac{SNR} points for Bob and Eve at which we acquire the desired input distribution in Sec. \ref{Sec:CCon}. The design \acp{SNR} for estimating $H\left(U^l_i | \bm{V}^l_i, \bm{Y}\right)$ and $H\left(U^l_i | \bm{V}^l_i, \bm{Z}\right)$ are chosen to be $\SNR_B^*$ and $\SNR_E^*$, respectively. At both the encoder and the decoder, we use an \ac{SCL} decoder of list size $L_e = 16$ and $L_d = 16$, respectively. An 8-bit \ac{CRC} code is attached at the end of the message as an outer code. The shaping bits are determined according to the argmax rule (\ref{eq:argmax_rule}). We set $\kappa_D$ to be $1.15$ for 4-ASK, $1.1$ for 8-ASK, and $1.05$ for 16-ASK.

Please note that many wiretap coding schemes \cite{klinc2011ldpc,Baldi2012coding,Zhang2014polar,Nooraiepour2017Rand,Pfeiffer2022design}, especially coded modulation schemes \cite{Matsumine2022Security, Pfeiffer2022multi}, are designed for \ac{BER}-based security. However, a high \ac{BER} at Eve can not necessarily derive a low leakage $I(\bm{M};\bm{Z})$ considered in our work, and it is difficult to estimate the information leakage for these codes themselves and their application to coded modulation systems. For a fair comparison, we extend the polar wiretap codes \cite{mahdavifar2011achieving} to high-order modulation by directly applying the \ac{MLC} construction as a baseline. This scheme, referred to as the uniform case hereafter, is in fact a special case of the proposed coding scheme when $\mathcal{D}=\emptyset$, and can be easily proved to achieve the secrecy rate of \ac{DMWC} with uniform constellation input.

\begin{figure}[!t]
    \centering
    \includegraphics[width=0.42\textwidth]{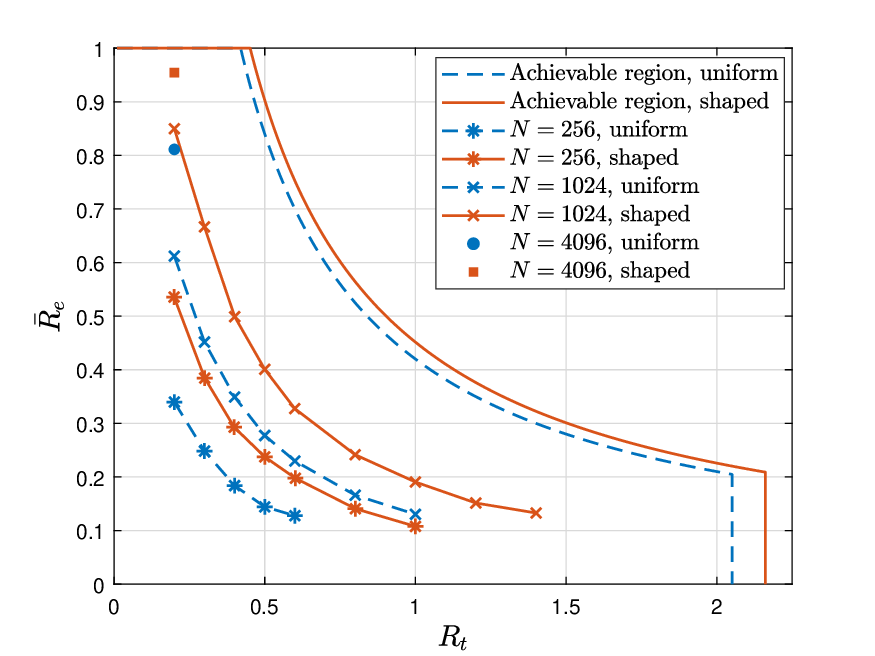}
    \caption{The achievable normalized rate-equivocation pairs for 8-ASK input over the Gaussian wiretap channel, where $P_{e,\max}^B = 10^{-5}$, $\SNR_B = 13$ dB, and $\SNR_E = 10$ dB.}
    \label{fig:Re_8ASK}
    \includegraphics[width=0.42\textwidth]{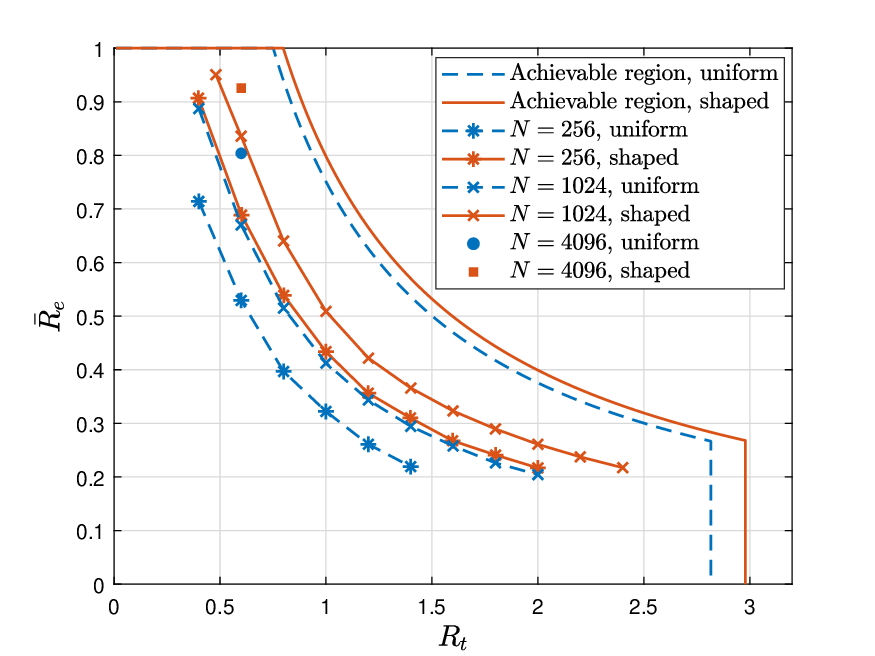}
    \caption{The achievable normalized rate-equivocation pairs for 16-ASK input over the Gaussian wiretap channel, where $P_{e,\max}^B = 10^{-5}$, $\SNR_B = 18$ dB, and $\SNR_E = 13$ dB.}
    \label{fig:Re_16ASK}
\end{figure}

Fig.~\ref{fig:simu_BER} illustrates the curves of Eve's \ac{BER} $P_e^E$ and the estimated normalized information leakage $L_k^E$ as $\SNR_B - \SNR_E$ increases for different transmission parameters over the Gaussian wiretap channel. The coding parameters are obtained according to $d\SNR_B = \SNR_B$, $d\SNR_g$ given in the figure, the security threshold $L_{k,\max}^E = 0.2$, and the reliability threshold $P_{e,\max}^B = 10^{-5}$. It can be observed that the proposed shaped scheme achieves higher $P_e^E$ and lower $L_k^E$ at eavesdropper compared to the uniform case, that is, achieves more secure communication. The security gap $S_g$ is calculated by (\ref{eq:Sg}) and also presented in Fig.~\ref{fig:simu_BER}. The results show that the proposed scheme achieves at least $0.6~\text{dB}$ gain on $S_g$. Furthermore, we observe that our leakage-based security condition ensures that Eve's \ac{BER} $P_e^E$ is close to 0.5, while the \ac{BER}-based security condition, with a typical value of $P_{e,\min}^E = 0.49$ for the security threshold \cite{klinc2011ldpc,Baldi2012coding,Zhang2014polar,Nooraiepour2017Rand,Pfeiffer2022design}, causes a normalized information leakage even larger than 0.6. This indicates that the estimated information leakage is a more stringent security metric than the \ac{BER} in terms of wiretap channel codes design.

If a security gap $S_g=3$ dB with the reliability threshold $P_{e,\max}^B = 10^{-5}$ and the security threshold $L_{k,\max}^E = 0.2$ is required, we find the maximum achievable message length $K$ for 4-\ac{ASK}, 8-\ac{ASK}, and 16-\ac{ASK} input at different $\SNR_B$ by numerical simulations with $d\SNR_B = \SNR_B$ and $d\SNR_g = S_g$. The results are illustrated in Fig.~\ref{fig:K3dB}. We observe that the shaped scheme always obtains a higher value of $K$ across $\SNR_B$ and modulation order $Q$. That is to say, under the same requirements on the reliability performance and security performance, the proposed shaped scheme is more bandwidth efficient than the uniform input case.

\begin{figure}[!t]
    \centering
    \includegraphics[width=0.42\textwidth]{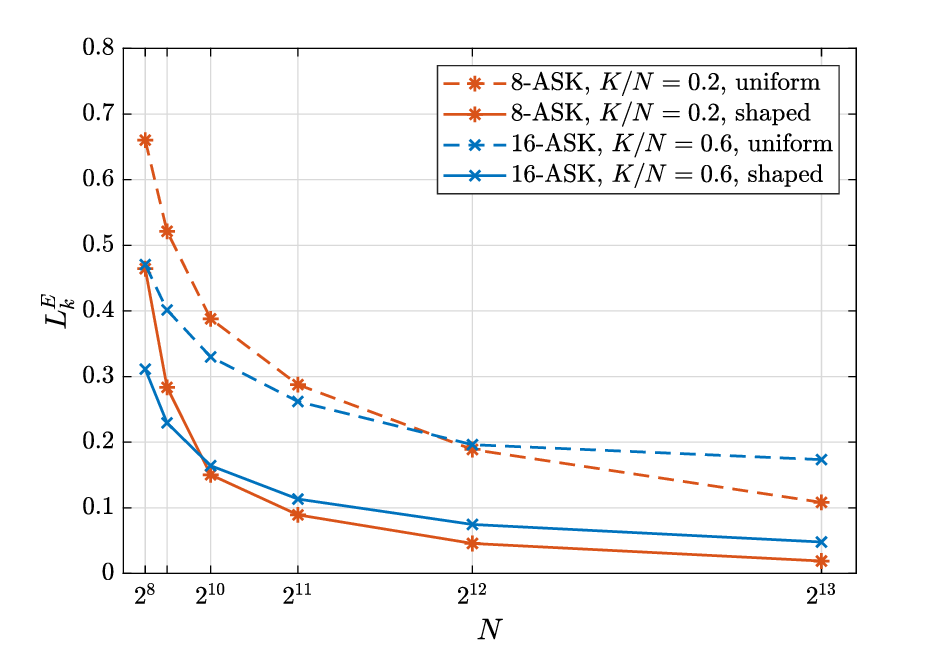}
    \caption{The estimated leakage $L_k^E$ as a function of $N$, where $P_{e,\max}^B = 10^{-5}$. For 8-ASK, $\SNR_B = 13$ dB and $\SNR_B = 10$ dB, while for 16-ASK, $\SNR_B = 18$ dB and $\SNR_B = 13$ dB.}
    \label{fig:LkN}
\end{figure}

\begin{figure*}[t]
    \centering
    \captionsetup[subfloat]{justification=centering}
    \subfloat[8-ASK, $K=400$, $\SNR_B=22$ dB, \\ and $d\SNR_g=10$ dB.]{
        \includegraphics[width=0.32\textwidth]{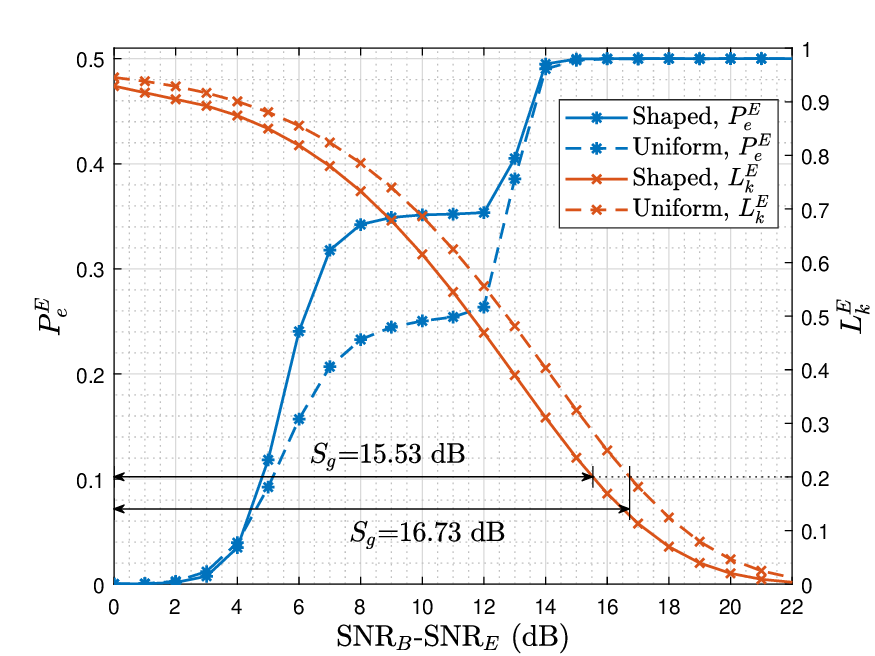}
    }\hspace{-0.2cm}
    \subfloat[16-ASK, $K=600$, $\SNR_B=32$ dB, \\ and $d\SNR_g=15$ dB.]{
        \includegraphics[width=0.32\textwidth]{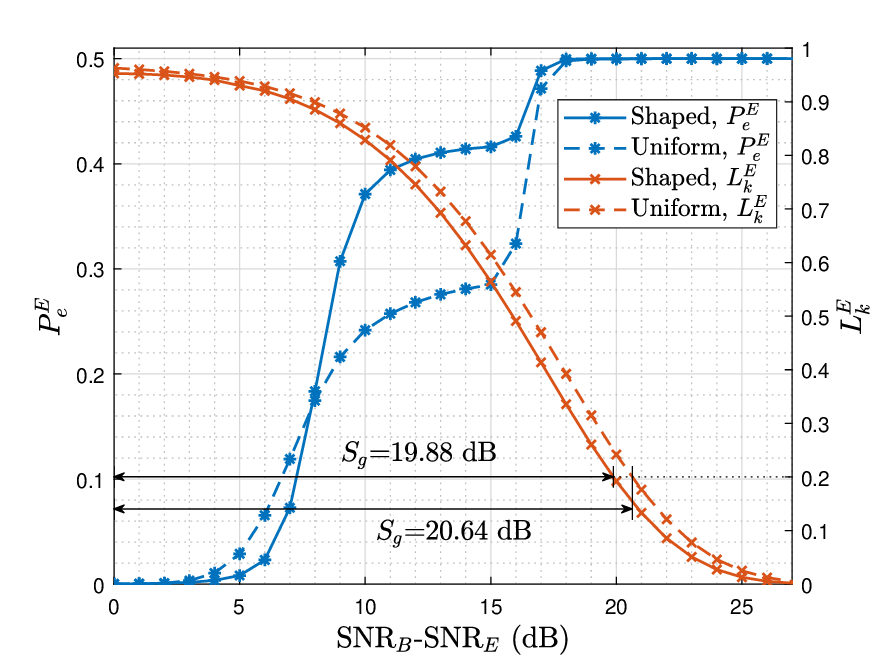}
    }\hspace{-0.2cm}
    \subfloat[64-QAM, $K=400$, $\SNR_B=22$ dB, \\ and $d\SNR_g=10$ dB.]{
        \includegraphics[width=0.32\textwidth]{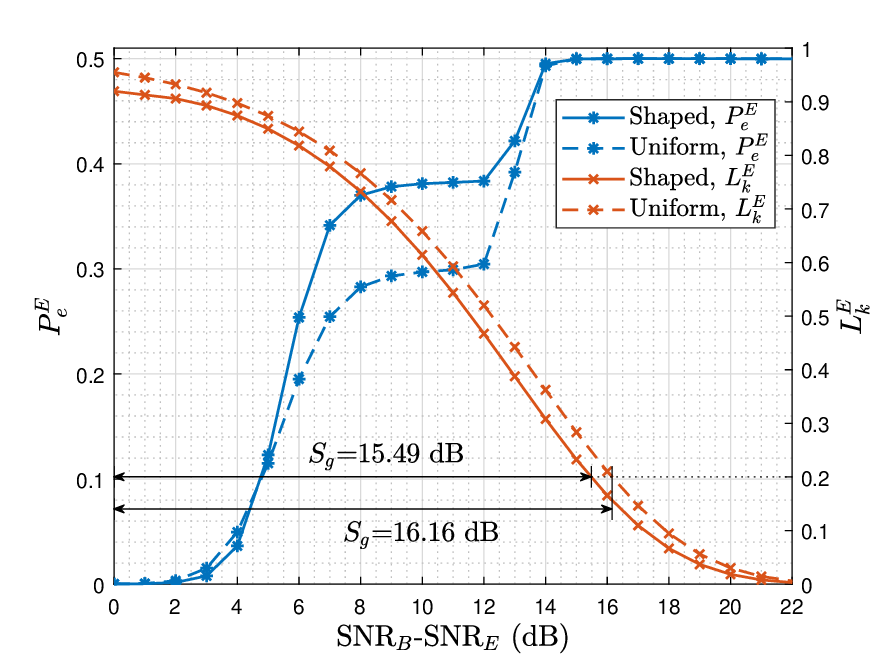}
    }
    \caption{Eve's BER $P_e^E$ and the estimated normalized information leakage $L_k^E$ versus $\SNR_B-\SNR_E$ for different modulation orders and message length $K$ over the \ac{RF} wiretap channel, where $L_{k,\max}^E = 0.2$, $P_{e,\max}^B = 10^{-4}$ and $N = 1024$.}
    \label{fig:simu_BER_F}
\end{figure*}

To further reveal the relationship between transmission efficiency and security, we present the normalized rate-equivocation pairs $(R_t, \bar{R}_e)$ achieved by the proposed shaped scheme and the uniform scheme at finite block length regimes, where the equivocation rate $R_e$ is normalized to $\bar{R}_e = R_e/R_t$ for better illustration. Note that since $\bar{R}_e$ can be expressed in the form of information leakage, i.e.,
$\bar{R}_e = H(\bm{M}|\bm{Z})/K = 1 - I(\bm{M};\bm{Z})/K$,
we can use the estimated normalized information leakage $L_k^E$ to compute $\bar{R}_e$. Now, Figs.~\ref{fig:Re_8ASK}~and~\ref{fig:Re_16ASK} depict the achievable normalized rate-equivocation pairs for 8-\ac{ASK} and 16-\ac{ASK} inputs with a reliability threshold $P_{e,\max}^B = 10^{-5}$ and different block length, respectively. The boundary of the achievable rate-equivocation region defined by (\ref{eq:Re_rg}) is also provided. We observe that there is a trade-off between transmission efficiency and security, that is, a higher transmission rate $R_t$ will cause a less normalized equivocation rate $\bar{R}_e$, indicating a higher risk of information leakage. Nevertheless, the curve generated by the pairs $(R_t, \bar{R}_e)$ with the same $N$ of the proposed shaped scheme always lies above the one obtained by the uniform scheme, which means that the proposed scheme is more bandwidth efficient and secure than the uniform input case. Specifically, the proposed scheme enables higher transmission rate when satisfying the same security condition, or equivalently, the proposed scheme achieves more secure transmission when satisfying the same reliability condition and transmission rate.

Moreover, as the block length $N$ increases, Figs.~\ref{fig:Re_8ASK} and \ref{fig:Re_16ASK} show that the pairs $(R_t, \bar{R}_e)$ asymptotically approach the boundary of the achievable rate-equivocation region (i.e., the limits for $N\to\infty$). In Fig.~\ref{fig:LkN}, we further show the estimated leakage $L_k^E$ as a function of $N$ for a given transmission rate, which indicates that the information leakage caused by our shaped scheme is obviously lower than the one of the uniform scheme, and tends to zero as $N\to\infty$. This also verifies that the proposed scheme satisfies the weak security condition, as proved in Theorem \ref{Thm:Secr}.

In addition, we assess the performance of our scheme over the \ac{RF} wiretap channel, where the gains of Bob's channel and Eve's channel follow the \ac{i.i.d.} complex Gaussian distribution with zero-mean and unit variance, respectively. We assume that Bob and Eve know their respective channel state information perfectly. Fig.~\ref{fig:simu_BER_F} illustrates Eve's \ac{BER} $P_e^E$ and the estimated leakage $L_k^E$ as $\SNR_B-\SNR_E$ increases over the \ac{RF} wiretap channel, which presents analogous performance to that observed over the Gaussian wiretap channel. If we choose the reliability threshold $P_{e,\max}^B = 10^{-4}$ and the security threshold $L_{k,\max}^E = 0.2$, the security gap $S_g$ is also provided in Fig.~\ref{fig:simu_BER_F}. We observe that 
fading brings a great challenge to secure transmission, reflected in the significant increase of the security gap. This is mainly because, due to the impact of fading, the instantaneous \ac{SNR} of Eve's channel may be superior to that of Bob's channel, which leads to the direct exposure of message to Eve. Nevertheless, the proposed shaped scheme still achieves a gain over 0.65 dB on security gap $S_g$ compared to the uniform input case.

\begin{table}[!t]
    \centering
    \fontsize{9pt}{10pt}\selectfont
    \newcolumntype{C}[1]{>{\centering\arraybackslash}p{#1}}
    \caption{Impact of Encoding List Size $L_e$ and Decoding List Size $L_d$ on the Security Gap $S_g$}
    \begin{tabular}{@{}C{1cm}C{1cm}C{1.6cm}C{1.6cm}@{}}
    \toprule
    \multirow{2}{*}{$L_e$} & \multirow{2}{*}{$L_d$} & \multicolumn{2}{c}{$S_g$} \\ \cmidrule(l){3-4} 
     & & Shaped & Uniform \\ \midrule
    16 & 16 & 3.90 dB & \multirow{2}{*}{4.54 dB} \\
    8  & 16 & 3.91 dB & \\
    8  & 8  & 3.99 dB & \multirow{2}{*}{4.70 dB} \\
    4  & 8  & 4.00 dB & \\
    4  & 4  & 4.19 dB & \multirow{2}{*}{4.94 dB} \\
    2  & 4  & 4.22 dB & \\
    \bottomrule
    \end{tabular}
    \label{tab:L_Sg}
\end{table}

Finally, we evaluate the impact of the list sizes for encoding and decoding on the security performance, as shown in \text{Table \ref{tab:L_Sg}}. Other simulation parameters except the list sizes are the same as that in Fig. \ref{fig:simu_BER}(a). Since $L_e$ and $L_d$ affect the shaping and decoding performance, respectively, an increase in $S_g$ is observed as they decrease. Nonetheless, the security gap $S_g$ is less sensitive to $L_e$ compared to $L_d$. The results tell us that there is a trade-off between the complexity and the security performance of the proposed scheme, and we need to determine the list sizes according to practical requirements.


\section{Conclusion}
In this work, we have performed \ac{PS} in the wiretap channel model. Under the constraints of Gaussian wiretap channel with \ac{ASK}/\ac{QAM} constellation input, we have found a sub-optimal solution to the input distribution optimization problem that maximizes the achievable secrecy rate $R_s$. The resulting sub-optimal solution $P_{X^*}$ can still lead to an $R_s$ close to the Gaussian secrecy capacity. In order to achieve secure communication with shaped input, we have proposed a probabilistic shaped multilevel polar coding scheme. Specifically, for the asymmetric \ac{DMWC} with constellation input, we have proven that the bit-channels of multilevel polar codes will polarize into four sets $\mathcal{F}$, $\mathcal{A}$, $\mathcal{R}$ and $\mathcal{D}$, and proposed to fill the bit positions indexed by these four sets with frozen bits, message bits, random bits and shaping bits, respectively. Such polar codes are proven to achieve the secrecy capacity of asymmetric \ac{DMWC}, while satisfying the reliability condition and the weak security condition. The above conclusions also hold for the Gaussian wiretap channel. Moreover, the coding procedure and security-oriented construction method of proposed multilevel polar codes for finite block length have been provided.

Numerical results have shown that the proposed scheme can achieve a more secure and bandwidth efficient transmission than the uniform input case over the Gaussian wiretap channel, that is, at least 0.6 dB gain on the security gap $S_g$ or significant improvement on the message length $K$. As block length $N$ tends to infinity, it has been verified that the proposed scheme satisfies the weak security condition. Furthermore, the proposed scheme also outperforms the uniform scheme over the \ac{RF} wiretap channel on the security gap. Future research may consider the shaped coded modulation scheme satisfying the strong security condition. Also, the security performance over the fading wiretap channel may be further improved by integrating with multiple-input multiple-output, precoding and other technologies.


\appendices

\section{Proof of Lemma \ref{Lem:DegWl}}
\label{Prof:Lem:DegWl}
By the definition of channel degradation, we know that there exists a \ac{DMC} $\tilde{W}$ such that $W_2\left(z | x \right) = \sum_{y} W_1 \left(y | x \right) \tilde{W} \left(z | y \right)$. By (\ref{eq:subch}) and invertibility of $f(\cdot)$, we have $P_{Z|\bm{B}} \left(z | \bm{b}\right) = \sum_{y} P_{Y|\bm{B}} \left(y | \bm{b}\right) P_{Z|Y} \left(z | y\right)$. Since the bit-levels are independent of each other, we further have
\begin{align}
W_2^l &\left(z,\bm{b}^{[\![l-1]\!]} | b^l\right) = P_{Z,\bm{B}^{[\![l-1]\!]}|B^l} \left(z,\bm{b}^{[\![l-1]\!]} | b^l\right) \notag\\
&= \sum_{\bm{b}^{[\![q]\!] \setminus [\![l]\!]}} P_{Z,\bm{B}^{[\![l-1]\!]}|\bm{B}^{[\![q]\!]\setminus[\![l-1]\!]}} \left(z,\bm{b}^{[\![l-1]\!]} | \bm{b}^{[\![q]\!]\setminus[\![l-1]\!]}\right) \notag\\ 
&= \sum_{\bm{b}^{[\![q]\!] \setminus [\![l]\!]}} P_{\bm{B}^{[\![l-1]\!]}|\bm{B}^{[\![q]\!]\setminus[\![l-1]\!]}}\left(\bm{b}^{[\![l-1]\!]} | \bm{b}^{[\![q]\!]\setminus[\![l-1]\!]}\right) \notag\\
&\qquad\qquad \cdot P_{Z|\bm{B}} \left(z | \bm{b}\right) \\
&= \sum_{y}\sum_{\bm{b}^{[\![q]\!] \setminus [\![l]\!]}} P_{\bm{B}^{[\![l-1]\!]}|\bm{B}^{[\![q]\!]\setminus[\![l-1]\!]}}\left(\bm{b}^{[\![l-1]\!]} | \bm{b}^{[\![q]\!]\setminus[\![l-1]\!]}\right) \notag\\
&\qquad\qquad \cdot P_{Y|\bm{B}} \left(y | \bm{b}\right) P_{Z|Y} \left(z | y\right) \notag\\
&= \sum_{y} W_1^l \left(y,\bm{b}^{[\![l-1]\!]} | b^l\right) \tilde{W} \left(z | y \right). \notag
\end{align}
Thus, we have $W_2^l \preceq W_1^l$.

\section{Proof of Lemma \ref{Lem:ZB}}
\label{Prof:Lem:ZB}
Let $\bm{V} = \left( \bm{S}, \bm{Y} \right)$ and $\bm{T} = \left( \bm{S}, \bm{Z} \right)$. Then, we have
\begin{align}
Z_B &\left(U_i | \bm{S}, \bm{Z} \right) \notag\\ 
& \overset{(a)}{=} 2 \sum_{\bm{t}} \sqrt{P_{U_i, \bm{T}} \left(0,  \bm{t} \right) P_{U_i, \bm{T}} \left( 1, \bm{t} \right)} \notag\\
& \overset{(b)}{=} 2 \sum_{\bm{t}} \sqrt{P_{U_i}(0) P_{\bm{T} | U_i} \left(\bm{t} | 0 \right) P_{U_i}(1) P_{\bm{T} | U_i} \left(\bm{t} | 1 \right)} \notag\\
& \overset{(c)}{=} 2 \sum_{\bm{t}} \sqrt{\sum_{\bm{v}} P_{U_i}(0) P_{\bm{V} | U_i} \left(\bm{v} | 0 \right) P_{\bm{T} | \bm{V}} \left(\bm{t} | \bm{v} \right) } \notag\\
& \qquad\quad \cdot \sqrt{\sum_{\bm{v}} P_{U_i}(1) P_{\bm{V} | U_i} \left(\bm{v} | 1 \right) P_{\bm{T} | \bm{V}} \left(\bm{t} | \bm{v} \right)} \\
& \overset{(d)}{\geq} 2 \sum_{\bm{t}} \sum_{\bm{v}} \sqrt{ P_{U_i}(0) P_{\bm{V} | U_i} \left(\bm{v} | 0 \right) P_{\bm{T} | \bm{V}} \left(\bm{t} | \bm{v} \right) } \notag\\
& \qquad\quad \cdot \sqrt{P_{U_i}(1) P_{\bm{V} | U_i} \left(\bm{v} | 1 \right) P_{\bm{T} | \bm{V}} \left(\bm{t} | \bm{v} \right)} \notag\allowdisplaybreaks \\
& = 2 \sum_{\bm{v}} \sqrt{P_{U_i, \bm{V}} \left(0,  \bm{v} \right) P_{U_i, \bm{V}} \left( 1, \bm{v} \right)} \sum_{\bm{t}} P_{\bm{T} | \bm{V}} \left(\bm{t} | \bm{v} \right) \notag\\
& = Z_B\left(U_i | \bm{S}, \bm{Y} \right). \notag
\end{align}
The equalities (a) and (b) are immediate from the definition of  Bhattacharyya parameter and conditional probability, respectively. By the definition of channel degradation, there exists a \ac{DMC} $\tilde{W}$ such that $\bar{W}_{2,N}\left(\bm{t} | u \right) = \sum_{\bm{v}} \bar{W}_{1,N} \left(\bm{v} | u \right) \tilde{W} \left(\bm{t} | \bm{v} \right)$. Combined with (\ref{eq:subch}), the equality (c) holds. Finally, the inequality (d) follows from the Cauchy-Schwartz inequality.

\section{Proof of Proposition \ref{Prop:HXY}}
\label{Prof:Prop:HXY}
For each bitlevel $l \in [\![q]\!]$, consider the sets
\begin{align}
    \mathcal{L}_{B^l} &= \left\{(l,i): Z_B\left(U_i^l| \bm{V}_i^l\right) < \delta_N \right\}, \\
    \mathcal{H}_{B^l} &= \left\{(l,i): Z_B\left(U_i^l| \bm{V}_i^l\right) > 1-\delta_N \right\}, \\
    \mathcal{L}_{B^l|Y} &= \left\{(l,i): Z_B\left(U_i^l| \bm{V}_i^l, \bm{Y}\right) < \delta_N \right\}, \\
    \mathcal{H}_{B^l|Y} &= \left\{(l,i): Z_B\left(U_i^l| \bm{V}_i^l, \bm{Y}\right) > 1-\delta_N \right\}.
\end{align}
By \cite[Lem. 1]{runge2022multilevel}, we have
\begin{equation}
    1 - \lim_{N \to \infty} \frac{\left| \mathcal{L}_{B^l} \right|}{N} = \lim_{N \to \infty} \frac{\left| \mathcal{H}_{B^l} \right|}{N} \quad = H(B^l|\bm{B}^{[\![l-1]\!]})
\end{equation}
and
\begin{equation}
    1 - \lim_{N \to \infty} \frac{\left| \mathcal{L}_{B^l|Y} \right|}{N} = \lim_{N \to \infty} \frac{\left| \mathcal{H}_{B^l|Y} \right|}{N} = H(B^l|\bm{B}^{[\![l-1]\!]},Y).
\end{equation}
Note that, for $l \neq k$, $\mathcal{L}_{B^l} \cap \mathcal{L}_{B^k} = \mathcal{H}_{B^l} \cap \mathcal{H}_{B^k} = \mathcal{L}_{B^l|Y} \cap \mathcal{L}_{B^k|Y} = \mathcal{H}_{B^l|Y} \cap \mathcal{H}_{B^k|Y} = \emptyset$ by definition. Thus
\begin{equation}
\begin{aligned}
    \mathcal{L}_X &= \bigcup_{l\in[\![q]\!]} \mathcal{L}_{B^l}, \quad~\mathcal{L}_{X|Y} = \bigcup_{l\in[\![q]\!]} \mathcal{L}_{B^l|Y}, \\
    \mathcal{H}_X &= \bigcup_{l\in[\![q]\!]} \mathcal{H}_{B^l}, \quad\mathcal{H}_{X|Y} = \bigcup_{l\in[\![q]\!]} \mathcal{H}_{B^l|Y}.
\end{aligned}
\end{equation}
~

Since $f(\cdot)$ is invertible, we have
\begin{align}
    \label{eq:profHX}
    \lim_{N \to \infty} \frac{1}{N} \big| \mathcal{L}_{X} \big| &= \lim_{N \to \infty} \frac{1}{N} \sum_{l\in[\![q]\!]} \big| \mathcal{L}_{B^l} \big| \notag\\ 
    &= \sum_{l\in[\![q]\!]} 1 - H(B^l|\bm{B}^{[\![l-1]\!]}) \\
    &= q - H(X). \notag
\end{align}
The proofs of the remaining equalities in (\ref{eq:NLHX}) and (\ref{eq:NLHXY}) follow analogously to (\ref{eq:profHX}).

\section{Proof of Theorem \ref{Thm:ML-FARD}}
\label{Prof:Thm:ML-FARD}
By (\ref{eq:NLHXY}), we immediately have
\begin{equation}
    \lim_{N \to \infty} \frac{\left|\mathcal{F}\right|}{N} = q - \lim_{N \to \infty} \frac{\left| \mathcal{L}_{X|Y} \right|}{N} = H(X|Y).
\end{equation}
From (\ref{eq:NLHX}), we know that $\lim_{N \to \infty} (\mathcal{H}_{X} \cup \mathcal{L}_{X})^C = \emptyset$. By the fact that $\mathcal{L}_{X} \subseteq \mathcal{L}_{X|Z} \subseteq \mathcal{L}_{X|Y}$, we have
\begin{equation}
\begin{aligned}
    \lim_{N \to \infty} \frac{\left|\mathcal{D}\right|}{N} &= \lim_{N \to \infty} \frac{\big| \mathcal{L}_{X|Y} \cap (\mathcal{L}_{X} \cup (\mathcal{H}_{X} \cup \mathcal{L}_{X})^C )\big|}{N} \\
    &= \lim_{N \to \infty} \frac{\big| \mathcal{L}_{X|Y} \cap \mathcal{L}_{X} \big|}{N} \\
    &= q-H(X).
\end{aligned}
\end{equation}
Additionally, from (\ref{eq:NLHXY}), we have
\begin{align}
    \lim_{N \to \infty} \frac{\left|\mathcal{A}\right|}{N} &= \lim_{N \to \infty} \frac{\big| (\mathcal{L}_{X|Y} \setminus \mathcal{L}_{X|Z}) \cap (\mathcal{L}_{X} \cup (\mathcal{H}_{X} \cup \mathcal{L}_{X})^C )\big|}{N} \notag\\
    &= \lim_{N \to \infty} \frac{\big| (\mathcal{L}_{X|Y} \setminus \mathcal{L}_{X|Z}) \cap \mathcal{L}_{X}\big|}{N} \notag\\
    &= \lim_{N \to \infty} \frac{\big| \mathcal{L}_{X|Y}\big| - \big| \mathcal{L}_{X|Z}\big|}{N} \\
    &= (q-H(X|Y))-(q-H(X|Z)) \notag\\
    &= I(X;Y) - I(X;Z) \notag
\end{align}
and
\begin{equation}
\begin{aligned}
    \lim_{N \to \infty} \frac{\left|\mathcal{R}\right|}{N} &= \lim_{N \to \infty} \frac{\big| \mathcal{L}_{X|Z} \setminus (\mathcal{L}_{X} \cup (\mathcal{H}_{X} \cup \mathcal{L}_{X})^C )\big|}{N} \\
    &= \lim_{N \to \infty} \frac{\big| \mathcal{L}_{X|Z} \setminus \mathcal{L}_{X} \big|}{N} \\
    &= (q-H(X|Z)) - (q-H(X)) \\
    &= I(X;Z). 
\end{aligned}
\end{equation}

\section{Proof of Lemma \ref{Lem:HRZA}}
\label{Prof:Lem:HRZA}
Suppose that Eve knows the realization $\bm{u}_\mathcal{A}$ of $\bm{U}_\mathcal{A}$ and wants to estimate $\bm{U}_\mathcal{R}$ with \ac{SC} decoding. Thus, Eve can treat both $\bm{U}_\mathcal{F}$ and $\bm{U}_\mathcal{A}$ as frozen bits. Since $\mathcal{R} = \mathcal{L}_{X|Z} \cap \mathcal{H}_{X}$, it is precisely the scenario considered in \cite{runge2022multilevel}. By \cite[Theorem 2]{runge2022multilevel}, the error probability of Eve's estimate $\hat{\bm{U}}_\mathcal{R}$ is given by
\begin{equation}
\label{eq:lambda}
    \lambda \overset{\text{def}}{=} \mathrm{Pr}\{\hat{\bm{U}}_\mathcal{R} \neq \bm{U}_\mathcal{R} \} \leq c2^{-N^{\beta^\prime}}
\end{equation}
for any $0<\beta^\prime<\beta<1/2$. Let $\bm{\mathcal{U}}_\mathcal{R}$ be the alphabet of $\bm{U}_\mathcal{R}$. Since $\bm{U}_\mathcal{R}$ is uniform over $\{0,1\}^{|\mathcal{R}|}$, the size of $\bm{\mathcal{U}}_\mathcal{R}$ is $2^{|\mathcal{R}|}$. Using Fano's inequality \cite[Theorem 2.10.1]{cover2006elements}, we bound the conditional entropy $H(\bm{U}_\mathcal{R}|\bm{Z},\bm{U}_\mathcal{A})$ as follows
\begin{equation}
\begin{aligned}
    H(\bm{U}_\mathcal{R}|\bm{Z},\bm{U}_\mathcal{A}) &\leq h_2(\lambda) + \lambda \log(|\bm{\mathcal{U}}_\mathcal{R}|) \\
    &\overset{(a)}{\leq} h_2(c2^{-N^{\beta^\prime}}) + c|\mathcal{R}|2^{-N^{\beta^\prime}}
\end{aligned}
\end{equation}
where (a) holds from (\ref{eq:lambda}), $|\bm{\mathcal{U}}_\mathcal{R}|=2^{|\mathcal{R}|}$, and the fact that the binary entropy function $h_2(\cdot)$ is an increasing function on the interval $[0, 1/2]$.

\section{Proof of Theorem \ref{Thm:Secr}}
\label{Prof:Thm:Secr}
The information leakage $I(\bm{M};\bm{Z})$ is upper bounded by
\begin{align}
\label{eq:IMZ}
    I(\bm{M};\bm{Z}) &\overset{(a)}{=} I(\bm{U}_\mathcal{A};\bm{Z}) \notag\\
    &\overset{(b)}{=} I(\bm{U}_{\mathcal{A} \cup \mathcal{F}};\bm{Z}) \notag\allowdisplaybreaks \\
    &\overset{(c)}{=} I(\bm{U};\bm{Z}) - I(\bm{U}_\mathcal{R};\bm{Z}|\bm{U}_{\mathcal{A} \cup \mathcal{F}}) \notag\\
    &\qquad - I(\bm{U}_\mathcal{D};\bm{Z}|\bm{U}_{\mathcal{A} \cup \mathcal{F} \cup \mathcal{R}}) \notag\\
    &\overset{(d)}{=} I(\bm{U};\bm{Z}) - I(\bm{U}_\mathcal{R};\bm{Z}|\bm{U}_{\mathcal{A} \cup \mathcal{F}}) \\
    &\overset{(e)}{=} I(\bm{U};\bm{Z}) - H(\bm{U}_\mathcal{R}|\bm{U}_\mathcal{A}) + H(\bm{U}_\mathcal{R}|\bm{Z},\bm{U}_\mathcal{A}) \notag\\
    &\overset{(f)}{=} I(\bm{U};\bm{Z}) - |\mathcal{R}| + H(\bm{U}_\mathcal{R}|\bm{Z},\bm{U}_\mathcal{A}) \notag\\
    &\overset{(g)}{\leq} N I(X;Z) - |\mathcal{R}| + H(\bm{U}_\mathcal{R}|\bm{Z},\bm{U}_\mathcal{A}) \notag\\
    &\overset{(h)}{\leq} N I(X;Z) - |\mathcal{R}| + h_2(c2^{-N^{\beta^\prime}}) + c|\mathcal{R}|2^{-N^{\beta^\prime}} . \notag
\end{align}
The equalities (a) and (b) follow from $\bm{U}_\mathcal{A} = \bm{M}$ and $\bm{U}_{\mathcal{F}}=\bm{0}$, respectively. Since the sets $\mathcal{F}$, $\mathcal{A}$, $\mathcal{R}$, $\mathcal{D}$ are disjoint and $\mathcal{F} \cup \mathcal{A} \cup \mathcal{R} \cup \mathcal{D} = \{(l,i): l\in [\![q]\!], i \in [\![N]\!]\}$, (c) holds by the chain rule of mutual information. Note that $\bm{U}_\mathcal{D}$ is deterministic given $\bm{U}_{\mathcal{A} \cup \mathcal{F} \cup \mathcal{R}}$. Hence, $I(\bm{U}_\mathcal{D};\bm{Z}|\bm{U}_{\mathcal{A} \cup \mathcal{F} \cup \mathcal{R}}) = 0$ and (d) holds. The equality (e) follows from $\bm{U}_{\mathcal{F}}=\bm{0}$ and the definition of conditional mutual information. Since $\bm{U}_\mathcal{R}$ and $\bm{U}_\mathcal{A}$ are independent, and $\bm{U}_\mathcal{R}$ is a priori uniform over $\{0,1\}^{|\mathcal{R}|}$, we obtain the equality (f). Inequality (g) follows from the fact that $\bm{U}$ is encoded to $\bm{X}$, and then $\bm{X}$ is transmitted over $N$ independent uses of $W_2$, where the achievable rate of $W_2$ is upper bound by $I(X;Z)$ for a given input distribution $P_X$. Finally, (h) is immediate from Lemma \ref{Lem:HRZA}.

Dividing both sides of (\ref{eq:IMZ}) by $N$, we have
\begin{equation}
\label{eq:IMZN}
    \frac{I(\bm{M};\bm{Z})}{N} = I(X;Z) - \frac{|\mathcal{R}|}{N} + \frac{h_2(c2^{-N^{\beta^\prime}})}{N} + \frac{c|\mathcal{R}|}{N} 2^{-N^{\beta^\prime}}.
\end{equation}
As $N \to\infty$, it is obvious that the last two terms in (\ref{eq:IMZN}) tend to 0 since $|\mathcal{R}| = \varTheta(N)$. Moreover, we have $\lim_{N \to\infty} I(X;Z) - |\mathcal{R}|/N = 0$ by (\ref{eq:NR_IXY}). Thus, we complete the proof of the theorem.

\bibliographystyle{bibdata/IEEEtran.bst}
\bibliography{bibdata/IEEEabrv,bibdata/refer}

\end{document}